\documentclass[aps,prl,twocolumn,nofootinbib]{revtex4}
\usepackage{graphicx}
\usepackage[utf8]{inputenc}
\usepackage{placeins}
\usepackage{amssymb, amsmath}
\usepackage{tabularx}
\usepackage[normalem]{ulem}

\newcommand{\Ms}{M_{\odot}}

\newcommand{\rns}{\rho_{\rm sat}}

\newcommand{\dops}{doppelg\"angers~}

\newcommand{\dop}{doppelg\"anger~}

\usepackage[dvipsnames, usenames]{xcolor}

\begin{document}

\title{Degeneracy in the inference of phase transitions in the neutron star equation of state from gravitational wave data}
\author{Carolyn A. Raithel}
\email{craithel@ias.edu}
\author{Elias R. Most}
\email{emost@princeton.edu}
\affiliation{School of Natural Sciences, Institute for Advanced Study, 1 Einstein Drive, Princeton, NJ 08540, USA}
\affiliation{Princeton Center for Theoretical Science, Jadwin Hall, Princeton University, Princeton, NJ 08544, USA}
\affiliation{Princeton Gravity Initiative, Jadwin Hall, Princeton University, Princeton, NJ 08544, USA}
\date{May 2021}

\begin{abstract}
 \noindent  Gravitational wave (GW) detections of binary neutron star inspirals
 will be crucial for constraining the dense matter equation of state (EoS).
 We demonstrate
 a new degeneracy in the mapping from tidal deformability
 data to the EoS, which occurs for models
 with strong phase transitions. We find that there exists a new family
 of EoS with phase transitions
 that set in at different densities and that predict
 neutron star radii that differ by up to $\sim500$~m, but that produce
 nearly identical tidal deformabilities for all neutron star masses. 
 Next generation GW detectors and advances in
 nuclear theory may be needed to resolve this degeneracy.
 
\end{abstract}

\maketitle

\textit{Introduction.---} Gravitational wave (GW) events offer exciting prospects to constrain the
properties of dense matter \cite[e.g.,][]{Agathos:2015uaa,Raithel:2019uzi,Baiotti:2019sew,Chatziioannou:2021tdi}.
In particular, the GW signal emitted
during the final orbits of two colliding neutron stars contains
imprints of the tidal deformability parameter, $\widetilde{\Lambda}$,
that can be related to the properties of dense matter 
in terms of the equation of state (EoS)
\cite[e.g.,][]{Annala:2017llu,Radice:2017lry,Bauswein:2017vtn,Most:2018hfd,LIGOScientific:2018cki,Raithel:2018ncd,De:2018uhw,Chatziioannou:2018vzf,Carson:2018xri}.
In practice, this inference is limited by the sensitivity to which
the tidal deformability can be constrained. For example, for the GW170817 event, 
$\widetilde{\Lambda}$ was constrained to $300\substack{+420\\-230}$
at 90\% confidence \cite{LIGOScientific:2018hze},
which has been translated to constraints on the neutron star radius of
$10 \lesssim R \lesssim 13$~km \cite[e.g.,][]{Annala:2017llu,Bauswein:2017vtn,De:2018uhw,Most:2018hfd,Raithel:2018ncd,LIGOScientific:2018cki}.

When advanced LIGO reaches its fifth observing campaign (called ``A+"), 
it is expected that the tidal deformability will be able to be constrained
to uncertainties of
$\sigma_{\widetilde{\Lambda}}\approx 46$ at 68\% confidence, for a 
GW170817-like event.
With next-generation (XG) GW detectors, 
these bounds on $\widetilde{\Lambda}$
will be further improved, leading to anticipated constraints of
$\sigma_{\widetilde{\Lambda}} < 8$ from the inspiral GW signal
for a similar event, and $\sigma_{\widetilde{\Lambda}}\approx1-4$ 
for a population of mergers observed with XG detectors, depending
on the merger rate
\cite{Carson:2019rjx}. 
From the usual quasi-universal
relations that map tidal deformabilities to the neutron star radius
\cite[e.g.,][]{Yagi:2015pkc,Yagi:2016bkt,De:2018uhw,Raithel:2018ncd,Zhao:2018nyf},
one would typically assume that small uncertainties
in $\sigma_{\widetilde{\Lambda}}$ directly translate to
tight constraints on $R$, potentially
to 50-200~m accuracy \cite{Chatziioannou:2021tdi},
assuming that dynamical tides are correctly accounted 
for in the extraction of $\widetilde{\Lambda}$ \cite{Pratten:2021pro,Gamba:2022mgx}.
In all of these efforts, a key goal is to determine what relevant degrees of freedom
exist in the dense-matter cores of neutron stars -- for example,
whether there exists a phase transition (e.g., to deconfined quark matter), what the nature of the phase transition
is, and at what densities this transition occurs \cite[e.g.,][]{Paschalidis:2017qmb,Christian:2018jyd,Dexheimer:2018dhb,Han:2018mtj,Sieniawska:2018zzj,Chatziioannou:2019yko,Tan:2021nat,Bogdanov:2022faf}. 

In this work, we identify a new degeneracy in the mapping
from the tidal deformability to the neutron star EoS,
which arises specifically for models with strong phase transitions at densities around nuclear saturation.
We demonstrate this degeneracy with an example Bayesian
inference of the EoS, using mock tidal deformability data generated from 
an EoS with a first-order phase transition. 
For the sensitivity of the A+ LIGO configuration, we find that
it will be difficult to differentiate between certain classes of EoSs that
have first-order phase transitions setting in at different densities, and that even models
with no phase transition at all can mimic the same tidal deformability
data. With the sensitivity of proposed XG detectors, the constraints
on the EoS remain broad, but the degeneracy between these different phase transitions
starts to resolve.

Using these inference results as motivation, we show 
that it is generically possible to construct EoS models that have
phase transitions that set in at significantly different densities --
leading to differences in the predicted stellar radii of
 $\sim300$~m  --
but that predict nearly identical
tidal deformabilities   
across the entire range of
astrophysically-observed neutron star masses (corresponding to, e.g., 
absolute differences of $\Delta \Lambda \lesssim5$ and fractional differences $\lesssim1-2\%$ for
intermediate-mass stars).  
Given the similarity of these models' macroscopic features, despite
their significant differences in underlying microphysics,
we refer to these EoSs as ``tidal deformability doppelg\"angers".

\begin{figure*} \centering
  \includegraphics[width=\textwidth]{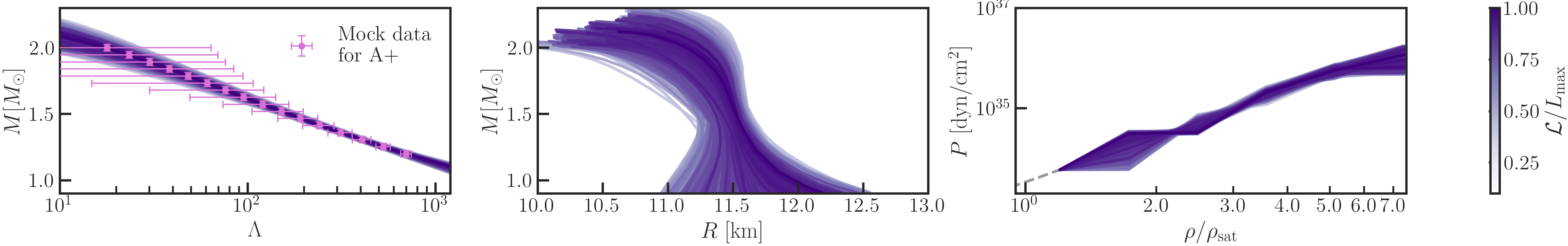}\\
    \includegraphics[width=\textwidth]{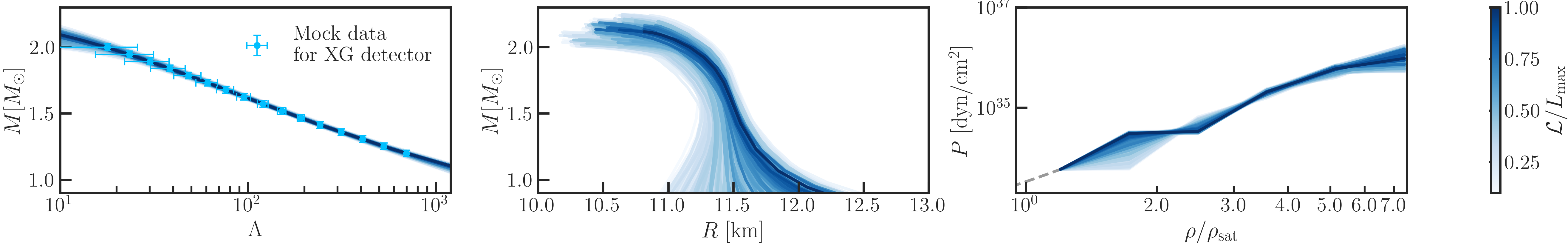}
  \caption{\label{fig:inference} Top row (in purple): Bayesian inference of the EoS for mock data, generated assuming Gaussian errors in tidal deformability, $\Lambda$, for LIGO at the sensitivity of its fifth observing run (A+), for a series of GW170817-like events ($\sigma_{\Lambda}$=46). From left to right, we show: the most likely tidal deformability curves, mass-radius curves, and EoSs inferred in our Bayesian inference. To highlight the degeneracy of the solutions, we randomly sample curves from the 68\% confidence interval and color them according to their posteriors, relative to the most likely solution. Bottom row (in blue): an identical inference, but with Gaussian errors in $\Lambda$ for the proposed XG detector Cosmic Explorer ($\sigma_{\Lambda}$=8).} 
\end{figure*}

We demonstrate that with additional input from nuclear theory
to extend the crust EoS to supranuclear densities
\cite[e.g.,][]{Gezerlis:2013ipa,Lynn:2015jua,Tews:2015ufa,Drischler:2017wtt},
combined with the sensitivity of XG GW observatories,
it may be possible to break the degeneracy when
inferring strong phase transitions. The degeneracy will be
easiest to break for low-mass neutron star binaries, if they exist in astrophysical populations.
At present, however, 
this degeneracy cannot be resolved from GW data alone,
and the inference of at least some families of phase transitions will be highly
sensitive to the choice of priors assumed.

\textit{Inferring strong phase transitions from GW data.---}
We begin by introducing the degeneracy with a sample Bayesian inference of the EoS
from mock GW data, following the statistical framework outlined in \cite{Raithel:2017ity}. In our inference scheme, we assume that the crust EoS (ap3 \cite{Akmal:1998cf}) is known perfectly to a fiducial density $\rho_0$, which we set here to be 1.2$\rns$, where $\rns=2.7\times10^{14}$~g/cm$^3$ is the nuclear saturation density.
At higher densities, we parameterize the uncertainty in the EoS using five piecewise polytropic segments. When performing our inference, we impose a set of minimal requirements: namely, causality, stability, and the ability to support massive ($2.01~\Ms$) neutron stars; and we sample uniformly in the pressure. Finally, we also require that the maximum mass predicted by the EoS not exceed 2.3~$\Ms$, in order to be consistent with inferences from GW170817 and its electromagnetic counterpart \cite{Margalit:2017dij,Rezzolla:2017aly,Ruiz:2017due,Shibata:2019ctb,Nathanail:2021tay}.

\begin{figure}[!ht]
\centering
\includegraphics[width=0.49\textwidth]{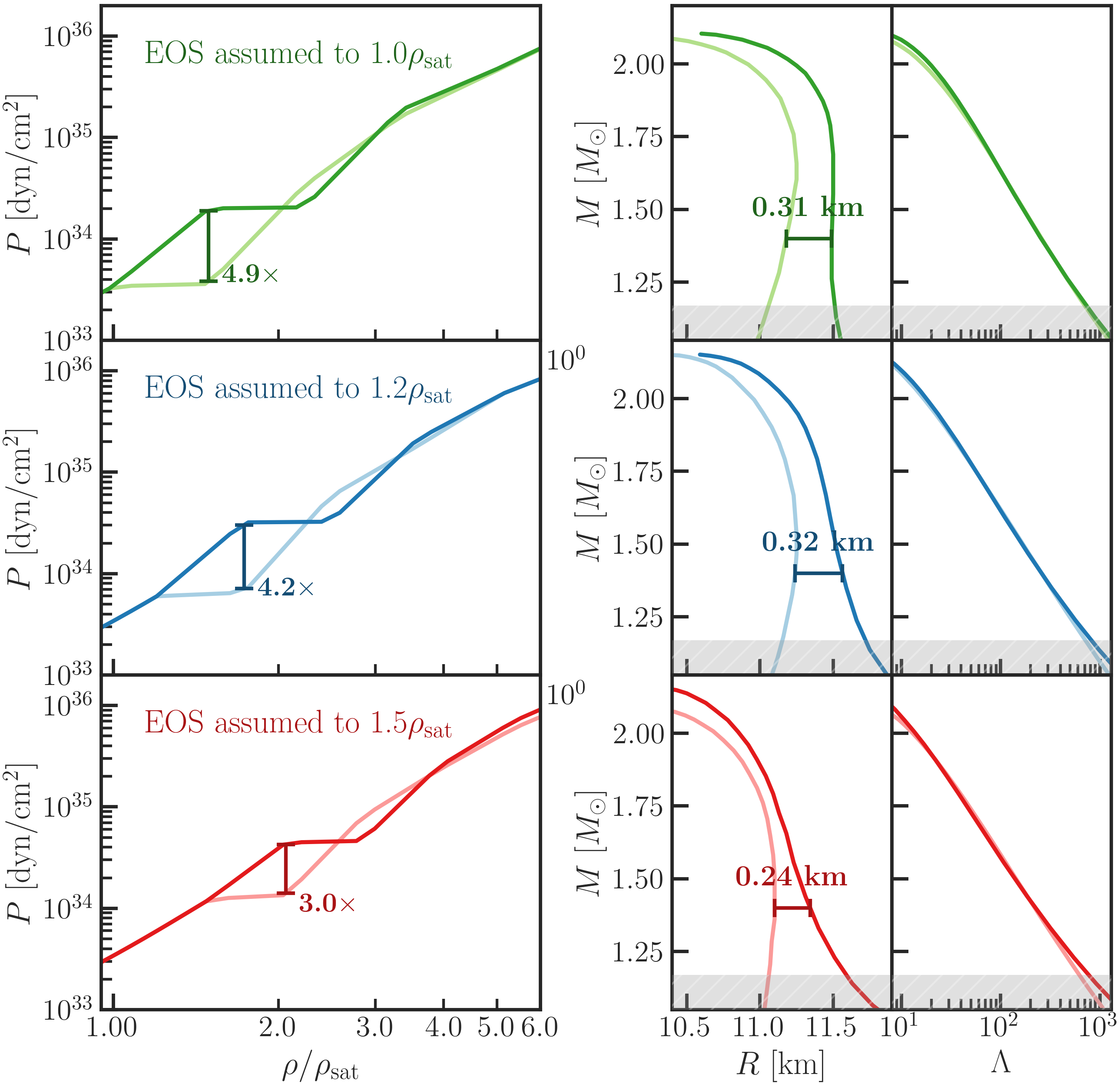}
\caption{\label{fig:dops} Example pairs of EoS models that undergo a first-order phase transition at significantly different densities 
(40\% fractional difference), and yet produce nearly identical tidal deformability curves.
From left to right, we show the EoS models, their corresponding mass-radius relations, and their corresponding tidal deformability curves. Each pair of EoS models was constructed assuming perfect knowledge of the crust EoS 
to $\rho_{\rm sat}$ (top row, in green), $1.2\rho_{\rm sat}$ (middle row, in blue), or $1.5\rho_{\rm sat}$ (bottom row, in red).} 
\end{figure}

For the example inferences, we construct a series of mock tidal deformability data
which were generated from an EoS that has a strong, first-order phase transition starting at 1.7$\rns$. This EoS
(which is shown as the dark blue curve in the middle row of Fig.~\ref{fig:dops}) predicts the radius
of a 1.4~$\Ms$ neutron star to be $R_{1.4}=11.6$~km and the tidal deformability at the same mass to be $\Lambda_{1.4}=257$,
consistent with current astrophysical constraints \cite{Ozel:2016oaf,Miller:2019cac,Riley:2019yda,Raaijmakers:2019qny,Raaijmakers:2019dks,Miller:2021qha,Riley:2021pdl,Raaijmakers:2021uju,LIGOScientific:2018cki,LIGOScientific:2018hze,Chatziioannou:2020pqz}.

In the first example inference, we assign Gaussian errors to the tidal deformabilities based on the projected sensitivity of the LIGO detectors in their fifth observing run (A+). We optimistically assume that the A+ detectors observe sixteen GW170817-like events, spaced evenly in mass across the entire range of astrophysically-observed neutron star masses (i.e., from 1.2-2$\Ms$).\footnote{We consider the range of astrophysical neutron star masses to range from the lightest observed radio pulsar at
1.17~$\Ms$ \cite{Martinez:2015mya}) to 2.01~$\Ms$
\cite{NANOGrav:2019jur,Fonseca:2021wxt}. Both quoted values
correspond to the 1-$\sigma$ lower limit on the masses. We
note that the 90\% lower-limit on the secondary mass for GW170817
was also 1.17$\Ms$, and that no lighter GW sources have yet been
detected \cite{LIGOScientific:2017vwq}.
In general, the differences in the tidal deformability are largest
at low masses, so taking the lower limit on the lowest mass considered provides
the most conservative estimates possible for the degeneracies discussed in this paper.} For such a scenario, the anticipated 1$\sigma$-measurement uncertainties in the tidal deformability would be $\sigma_{\Lambda}=46$ \cite{Carson:2019rjx}. We additionally assume that the masses are tightly constrained with Gaussian uncertainties of 0.025$\Ms$.\footnote{For simplicity, we also assume that these are equal mass binaries, so that the component tidal deformabilities are constrained directly. We explore the impact of unequal mass ratios on the degeneracy between these tidal deformabilities in the Supplemental Material.}

We show the resulting constraints in the top row of Fig.~\ref{fig:inference} (in purple) . In this figure, we include only samples drawn from the 68\% confidence interval for visual clarity, and we color these according to their normalized posteriors. We find that even with this optimistic set of mock data, we are only able to constrain the radius of a 1.4~$\Ms$ neutron star to to within 500~m and the pressure at 1.7$\rns$ to within a factor of 5.8$\times$, at 68\% confidence. Moreover, as the color-shading indicates, there are models on either edge of this broad uncertainty band that give comparably good posteriors. For example, models with a strong phase transition that sets in at lower densities (1.2$\rns$) or higher densities (1.7$\rns$) fit the data comparably well, as does an EoS that goes right through the middle of this uncertainty band with no phase transition at all. In short, we are not only unable to recover our initial EoS, but are also unable to rule out or confirm the presence of exotic nuclear phases, such as deconfined quark matter.

To further illustrate this degeneracy, we consider two example EoSs drawn from opposite edges of this uncertainty band. We show these EoSs in the middle row of Fig.~\ref{fig:dops}, where the dark blue curve shows the EoS model used to generate the mock data, and the light blue curve represents a separate EoS sampled in our inference. Despite the fact that these models predict first-order phase transitions at different densities and accordingly differ significantly in their supranuclear pressures, they fit the data similarly well with a Bayes factor of 1.9, indicating insufficient evidence to tell them apart \cite{Jeffreys61}.

This presents a significant degeneracy in the mapping from tidal deformability measurements to the underlying EoS. It has previously been shown that changing the crust EoS (at densities below $10^{14}$~g/cm$^3$) can change the radius without significantly affecting the tidal deformabilities \cite{Gamba:2019kwu}. Here, however, we find that large differences in the EoS at \textit{supranuclear} densities may also be indistinguishable, even with optimistic A+ data observed across a wide range of masses.

In order to understand the sensitivity of this degeneracy to the measurement uncertainties, we perform a second Bayesian inference with an identical set-up, 
but now assuming Gaussian errors on the tidal deformability measurements of $\sigma_{\Lambda}=8$. These smaller errors on $\Lambda$ correspond to the projected measurement uncertainty for the proposed XG detector Cosmic Explorer \cite{Reitze:2019iox}, for a GW170817-like event \cite{Carson:2019rjx}. 
The results of this inference are shown in the bottom row of Fig.~\ref{fig:inference} (in blue) . Again, we find a large spread in the inferred $R_{1.4}$ of $\sim500$~m and in the pressure at 1.7$\rns$ of 5.7$\times$, at 68\% confidence. However, in this case, we see that the data have more discerning power for the most extreme EoS models in our inferred sample, as indicated by the gradient in colors.

To illustrate this point, we again consider the two example EoSs from the middle row of Fig.~\ref{fig:dops}, which now have a Bayes factor of 3.3 for the XG data, indicating ``substantial" evidence \cite{Jeffreys61} in favor of the dark blue model (which was used to generate the mock data). Interestingly, the lowest-mass data points are the most constraining:  for example, if we excise the mock data point at 1.2~$\Ms$, then the Bayes factor between these models for the remaining data is only 1.6, which is insufficient evidence to select the correct EoS. This suggests that, if low-mass neutron star binaries exist in nature, they may be particularly powerful for resolving this degeneracy. We discuss this point further below.

In principle, with even one year of observations with Cosmic Explorer, we can expect tighter constraints on the tidal deformability than assumed here, of potentially $\sigma_{\Lambda}\approx1-4$, depending on the astrophysical merger rate \cite{Carson:2019rjx}. With such sensitivity -- in particular, if combined with further input from nuclear theory --  it will become possible to distinguish between these EoS models with higher confidence, as we show in the following section. 

Finally, we note that we have 
taken very broad priors in this analysis (namely, flat priors on the pressure with minimal additional physical constraints). We make this choice in order to clearly demonstrate the constraining power of these mock GW data. Additional priors on the sound speed or on the likelihood of phase transitions in the EoS would readily differentiate between the inferred EoSs shown in Fig.~\ref{fig:inference}. Indeed, we performed an additional set of inferences with more informative priors, and we find that the inclusion of even a weak prior penalizing variations in the sound speed acts to restrict the uncertainty bands in Fig.~\ref{fig:inference}, but that such a prior can also bias the inference to select the incorrect EoS, even in the limit of high-quality XG data observed across a range of masses (for details, see the Supplemental Material). This suggests a strong sensitivity of such inferences to the choice of priors, for at least some regions of the EoS parameter space where the \dop degeneracy is significant. In summary, these results demonstrate -- for the first time -- the limitations of current GW data in distinguishing between certain classes of EoS models with strong, supranuclear phase transitions, from the data directly. 

\textit{Impact on future gravitational wave detections.---}
In order to understand this degeneracy 
in more detail, we construct several example pairs of EoS models
that mimic the features identified in Fig.~\ref{fig:inference}. 
Because of their almost identical tidal deformabilities despite large
differences in their EoSs, we refer to these models as {\it tidal deformability
\dops}.

We show these ``\dop" EoSs in Fig.~\ref{fig:dops}.
The top row (in green) shows a pair of EoSs where the crust 
EoS is assumed to be known perfectly up to $\rns$, and the 
phase transition is allowed to set in soon thereafter. 
In the middle row (in blue), we show an example
where the crust EoS is assumed to 1.2$\rns$; these models correspond
to the extreme edges of our 68\% confidence band inferred in Fig.~\ref{fig:inference},
as discussed in that section.
Finally, in the bottom row (in red), we show an example pair of \dops
where the crust EoS is assumed to be known to 1.5$\rns$. In all cases, we
find that it is possible to construct pairs of EoS models
that have very different microphysics -- with first-order
phase transitions that set in at significantly different densities,
and which accordingly predict neutron star radii that differ by $\sim$300~m --
and yet that predict tidal deformability curves that are nearly identical
across the entire range of astrophysically observed masses.

We note that, although these are phenomenological models,
the qualitative features are similar to more realistic calculations
of EoSs with first-order phase transitions to deconfined quark matter
\cite[e.g.,][and references therein]{Kojo:2014rca,Baym:2017whm,Blaschke:2018mqw}. 
The difference in the transition densities
in Fig.~\ref{fig:dops} can thus be associated with a difference
in the deconfinement transition densities for these models; 
or, more generally, with the onset densities for an
exotic new degree of freedom.

\begin{figure}[!ht]
\centering
\includegraphics[width=0.49\textwidth]{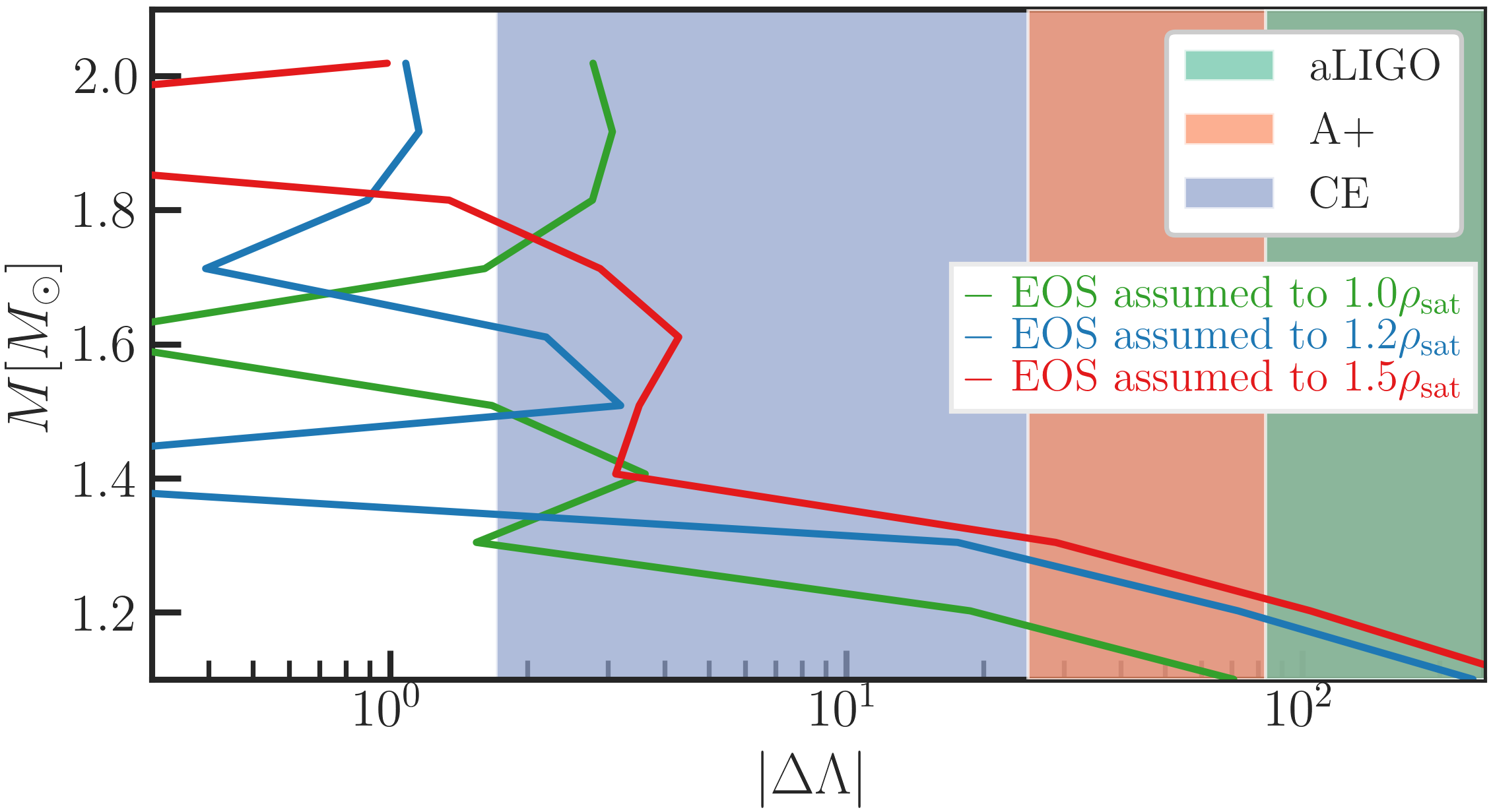}
\caption{\label{fig:dL} Absolute differences in the
tidal deformability,  $\Lambda$, between each pair of \dops shown 
in Fig.~\ref{fig:dops}. As the crust EoS is assumed to higher densities, 
the tidal deformability curves become more distinct, with
the largest differences emerging at low masses.
The vertical shaded bands indicate 
the expected 68\%-measurement uncertainty in $\Lambda$ for a 
population of neutron star
  mergers observed over one year with
  the sensitivity of LIGO at design sensitivity (aLIGO), 
  the anticipated sensitivity of LIGO during its fifth observing run (A+),
  and the proposed XG detector Cosmic Explorer (CE)
  \cite{Carson:2019rjx}. }
\end{figure}

Although the tidal deformability curves in  Fig.~\ref{fig:dops} 
are very similar for a given pair of models, they are not perfectly identical.
We show the differences in $\Lambda$ for each of these pairs of models in Fig.~\ref{fig:dL}, 
where, for reference,
we also include estimates of the differences in $\Lambda$ that could be
resolved at 68\% confidence for a population of neutron star mergers
observed for one year with the sensitivity of current and upcoming detectors. These
sensitivity estimates are shown with the vertical green band for aLIGO,
in orange for A+, and in blue for Cosmic Explorer \cite{Carson:2019rjx}.

We find that, in general, the differences in $\Lambda$ for any of these
EoSs are most significant at low masses, consistent with the findings
of the previous section. In addition, as the crust
EoS is assumed to higher densities, the differences in $\Lambda$ become more significant. 

 For example, if the crust EoS is assumed to be known to $\rns$ (1.5$\rns$),
we will likely need the sensitivity of Cosmic Explorer (A+, for a population of low-mass 
neutron star binaries) to distinguish the tidal deformabilities for
this example, based on GW data alone.
Thus, adopting stronger nuclear input -- in terms of the density to
which the crust EoS is assumed -- can help to resolve this tidal deformability degeneracy.
In summary, the most constraining data will likely come from low-mass 
neutron star binaries, which may even be able to resolve this degeneracy with
current GW detectors, if combined
with sufficient input from nuclear theory at supranuclear densities. 

We note that our discussion here focuses on a few 
illustrative examples, in order to discuss the implications
for inferences from current and upcoming observations. 
We investigate the ubiquity of these \dops and their full
parameter space in a separate work \cite{Raithel:2022aee}.

\textit{Prospects for post-merger GWs.---} 
In addition to providing tighter constraints on the tidal deformability
of inspiraling neutron stars,
another exciting prospect of XG detectors is the possibility of capturing
the post-merger GW emission 
  \cite[see, e.g.,][]{Wijngaarden:2022sah,Breschi:2022ens,Yu:2022upc}. Much work has been devoted to
understanding the connection of EoS models to the post-merger frequency
spectrum.
These quasi-universal relations rely to a large extent on
correlations between the dominant frequency $f_2$ and the tidal
deformabilities or radii of cold neutron stars 
\cite[e.g.,][]{Bauswein:2012ya,Takami:2014tva,Takami:2014zpa,Baiotti:2016qnr,Paschalidis:2016vmz,Rezzolla:2016nxn,Bauswein:2019ybt,Bernuzzi:2020tgt,Radice:2020ddv,Vretinaris:2019spn,Raithel:2022orm}.
In particular, several works have investigated the 
possibility of constraining strong phase transitions this way \cite{Most:2018eaw,Bauswein:2018bma,Weih:2019xvw,Most:2019onn,Prakash:2021wpz,Kedia:2022nns,Huang:2022mqp}.

To investigate this scenario,
we perform binary neutron star merger simulations for two extreme
pairs of \dop EoSs, where the crust EoS is assumed only to 0.5$\rns$.
In one pair of \dop EoSs, the characteristic radii are $R_{1.4}$=10.8 and 11.2~km,
similar to the examples shown in Fig.~\ref{fig:dops}.
We also construct a second pair of \dops
that are significantly stiffer, such that they predict characteristic radii of $R_{1.4}=12.8$
and 13.2~km, yet differ by $\Delta\Lambda_{1.4}<1$. We extend these zero-temperature, 
EoSs to finite-temperatures and 
arbitrary compositions using the framework of \cite{Raithel:2019gws}, and perform merger simulations for
each EoS using GW170817-like binary parameters. Two of these fully-finite
temperature models have been simulated previously
\cite{Most:2021ktk,Raithel:2022orm}, and the numerical set-up
\cite{Most:2019kfe,Etienne:2015cea} of our simulations here is identical to
that work \cite{Most:2021ktk}; we provide key details in the Supplemental Material.

From these simulations, we extract the peak frequencies
of the post-merger GW emission and find that they are nearly
indistinguishable for a given pair of \dops. For the $R_{1.4}=10.8$ 
and 11.2~km pair of models, we find $f_2=3.39$~kHz
in both cases; while for the $R_{1.4}=12.8$ and 13.2~km pair of models,
$f_2=2.71$ and 2.65~kHz, respectively.

These results are consistent with the predictions of
existing quasi-universal relations, to within the numerical 
uncertainties in $f_2$, which we conservatively estimate to be 
at the 10\% level \cite{Breschi:2019srl}. In particular, the \dops
do not violate reported quasi-universal relations between $f_2$ and 
the radius of a $1.8 M_\odot$ star \cite{Vretinaris:2019spn,Raithel:2022orm}, or with the tidal deformability \cite{Breschi:2022xnc}. For additional discussion, see the Supplemental Material. 

In summary, to within the current uncertainties of numerical simulations -- which may also be affected by systematic uncertainties in finite-temperature
\cite{Bauswein:2010dn,Figura:2020fkj,Raithel:2021hye} and neutrino physics
\cite{Alford:2017rxf,Most:2021zvc,Radice:2021jtw,Most:2022yhe} --, we find that the
post-merger peak frequencies may not be able to
differentiate between the strong phase transitions of some \dop models. However, the field is likely to progress significantly by the XG era.

\textit{Summary.---} 
In this work, we have identified a new degeneracy in the
mapping from tidal deformability data to the underlying EoS,
which arises for models with strong phase transitions.
We find that certain families of EoS models, which have
phase transitions that set in at significantly different
densities and which predict radii that differ by $\sim300$~m, 
can predict tidal deformabilities that are nearly
identical across the observed range of neutron star masses.

While this degeneracy may limit the ability of the current GW detectors
to infer some classes of phase transitions from GW data in the absence of 
informative priors; we have shown that XG detectors will potentially have the sensitivity
to resolve this degeneracy, depending on the neutron star mass distribution
and merger rate. These results thus provide additional motivation for
the construction of XG facilities such as Einstein Telescope
\cite{Punturo:2010zz}, Cosmic Explorer \cite{Reitze:2019iox}, or NEMO
\cite{Ackley:2020atn}.

Adopting stronger input from nuclear theory can also help to resolve
the degeneracy between certain classes of these models. Thus, continued
advances in
nuclear theoretical constraints -- in particular around nuclear
saturation \cite{Drischler:2021kxf,Tews:2022yfb} -- will also help to provide further
constraints on these tidal deformability \dops.
\\

\begin{acknowledgments}
\noindent The authors thank Gabriele Bozzola, Katerina Chatziioannou, Pierre Christian, Phil Landry,
Feryal \"Ozel, Dimitrios Psaltis, Jocelyn Read, Ingo Tews, and Nicolas Yunes 
for insightful comments on this work.
The authors gratefully acknowledge support from postdoctoral
fellowships at the Princeton Center for Theoretical Science, the Princeton
Gravity Initiative, and the Institute for Advanced Study.  CAR additionally
acknowledges support as a John N. Bahcall Fellow at the Institute for
Advanced Study.  This work was performed in part at the Aspen Center for
Physics, which is supported by National Science Foundation grant
PHY-1607611.  ERM acknowledges support for compute time allocations on the
NSF Frontera supercomputer under grants AST21006. This work used the
Extreme Science and Engineering Discovery Environment (XSEDE)
\cite{Towns:2014qtb} through Expanse at SDSC and Bridges-2 at PSC through
allocations PHY210053 and PHY210074. The simulations were also in part
performed on computational resources managed and supported by Princeton
Research Computing, a consortium of groups including the Princeton
Institute for Computational Science and Engineering (PICSciE) and the
Office of Information Technology's High Performance Computing Center and
Visualization Laboratory at Princeton University. The authors also
acknowledge the use of high-performance computing at the Institute for
Advanced Study.
\end{acknowledgments}

\bibliography{inspire,non_inspire,inspire_supplement}

\begin{thebibliography}{115}
\expandafter\ifx\csname natexlab\endcsname\relax\def\natexlab#1{#1}\fi
\expandafter\ifx\csname bibnamefont\endcsname\relax
  \def\bibnamefont#1{#1}\fi
\expandafter\ifx\csname bibfnamefont\endcsname\relax
  \def\bibfnamefont#1{#1}\fi
\expandafter\ifx\csname citenamefont\endcsname\relax
  \def\citenamefont#1{#1}\fi
\expandafter\ifx\csname url\endcsname\relax
  \def\url#1{\texttt{#1}}\fi
\expandafter\ifx\csname urlprefix\endcsname\relax\def\urlprefix{URL }\fi
\providecommand{\bibinfo}[2]{#2}
\providecommand{\eprint}[2][]{\url{#2}}

\bibitem[{\citenamefont{Agathos et~al.}(2015)\citenamefont{Agathos, Meidam,
  Del~Pozzo, Li, Tompitak, Veitch, Vitale, and Van
  Den~Broeck}}]{Agathos:2015uaa}
\bibinfo{author}{\bibfnamefont{M.}~\bibnamefont{Agathos}},
  \bibinfo{author}{\bibfnamefont{J.}~\bibnamefont{Meidam}},
  \bibinfo{author}{\bibfnamefont{W.}~\bibnamefont{Del~Pozzo}},
  \bibinfo{author}{\bibfnamefont{T.~G.~F.} \bibnamefont{Li}},
  \bibinfo{author}{\bibfnamefont{M.}~\bibnamefont{Tompitak}},
  \bibinfo{author}{\bibfnamefont{J.}~\bibnamefont{Veitch}},
  \bibinfo{author}{\bibfnamefont{S.}~\bibnamefont{Vitale}}, \bibnamefont{and}
  \bibinfo{author}{\bibfnamefont{C.}~\bibnamefont{Van Den~Broeck}},
  \bibinfo{journal}{Phys. Rev. D} \textbf{\bibinfo{volume}{92}},
  \bibinfo{pages}{023012} (\bibinfo{year}{2015}), \eprint{1503.05405}.

\bibitem[{\citenamefont{Raithel}(2019)}]{Raithel:2019uzi}
\bibinfo{author}{\bibfnamefont{C.~A.} \bibnamefont{Raithel}},
  \bibinfo{journal}{Eur. Phys. J. A} \textbf{\bibinfo{volume}{55}},
  \bibinfo{pages}{80} (\bibinfo{year}{2019}), \eprint{1904.10002}.

\bibitem[{\citenamefont{Baiotti}(2019)}]{Baiotti:2019sew}
\bibinfo{author}{\bibfnamefont{L.}~\bibnamefont{Baiotti}},
  \bibinfo{journal}{Prog. Part. Nucl. Phys.} \textbf{\bibinfo{volume}{109}},
  \bibinfo{pages}{103714} (\bibinfo{year}{2019}), \eprint{1907.08534}.

\bibitem[{\citenamefont{Chatziioannou}(2022)}]{Chatziioannou:2021tdi}
\bibinfo{author}{\bibfnamefont{K.}~\bibnamefont{Chatziioannou}},
  \bibinfo{journal}{Phys. Rev. D} \textbf{\bibinfo{volume}{105}},
  \bibinfo{pages}{084021} (\bibinfo{year}{2022}), \eprint{2108.12368}.

\bibitem[{\citenamefont{Annala et~al.}(2018)\citenamefont{Annala, Gorda,
  Kurkela, and Vuorinen}}]{Annala:2017llu}
\bibinfo{author}{\bibfnamefont{E.}~\bibnamefont{Annala}},
  \bibinfo{author}{\bibfnamefont{T.}~\bibnamefont{Gorda}},
  \bibinfo{author}{\bibfnamefont{A.}~\bibnamefont{Kurkela}}, \bibnamefont{and}
  \bibinfo{author}{\bibfnamefont{A.}~\bibnamefont{Vuorinen}},
  \bibinfo{journal}{Phys. Rev. Lett.} \textbf{\bibinfo{volume}{120}},
  \bibinfo{pages}{172703} (\bibinfo{year}{2018}), \eprint{1711.02644}.

\bibitem[{\citenamefont{Radice et~al.}(2018)\citenamefont{Radice, Perego,
  Zappa, and Bernuzzi}}]{Radice:2017lry}
\bibinfo{author}{\bibfnamefont{D.}~\bibnamefont{Radice}},
  \bibinfo{author}{\bibfnamefont{A.}~\bibnamefont{Perego}},
  \bibinfo{author}{\bibfnamefont{F.}~\bibnamefont{Zappa}}, \bibnamefont{and}
  \bibinfo{author}{\bibfnamefont{S.}~\bibnamefont{Bernuzzi}},
  \bibinfo{journal}{Astrophys. J. Lett.} \textbf{\bibinfo{volume}{852}},
  \bibinfo{pages}{L29} (\bibinfo{year}{2018}), \eprint{1711.03647}.

\bibitem[{\citenamefont{Bauswein et~al.}(2017)\citenamefont{Bauswein, Just,
  Janka, and Stergioulas}}]{Bauswein:2017vtn}
\bibinfo{author}{\bibfnamefont{A.}~\bibnamefont{Bauswein}},
  \bibinfo{author}{\bibfnamefont{O.}~\bibnamefont{Just}},
  \bibinfo{author}{\bibfnamefont{H.-T.} \bibnamefont{Janka}}, \bibnamefont{and}
  \bibinfo{author}{\bibfnamefont{N.}~\bibnamefont{Stergioulas}},
  \bibinfo{journal}{Astrophys. J. Lett.} \textbf{\bibinfo{volume}{850}},
  \bibinfo{pages}{L34} (\bibinfo{year}{2017}), \eprint{1710.06843}.

\bibitem[{\citenamefont{Most et~al.}(2018)\citenamefont{Most, Weih, Rezzolla,
  and Schaffner-Bielich}}]{Most:2018hfd}
\bibinfo{author}{\bibfnamefont{E.~R.} \bibnamefont{Most}},
  \bibinfo{author}{\bibfnamefont{L.~R.} \bibnamefont{Weih}},
  \bibinfo{author}{\bibfnamefont{L.}~\bibnamefont{Rezzolla}}, \bibnamefont{and}
  \bibinfo{author}{\bibfnamefont{J.}~\bibnamefont{Schaffner-Bielich}},
  \bibinfo{journal}{Phys. Rev. Lett.} \textbf{\bibinfo{volume}{120}},
  \bibinfo{pages}{261103} (\bibinfo{year}{2018}), \eprint{1803.00549}.

\bibitem[{\citenamefont{Abbott et~al.}(2018)}]{LIGOScientific:2018cki}
\bibinfo{author}{\bibfnamefont{B.~P.} \bibnamefont{Abbott}}
  \bibnamefont{et~al.} (\bibinfo{collaboration}{LIGO Scientific, Virgo}),
  \bibinfo{journal}{Phys. Rev. Lett.} \textbf{\bibinfo{volume}{121}},
  \bibinfo{pages}{161101} (\bibinfo{year}{2018}), \eprint{1805.11581}.

\bibitem[{\citenamefont{Raithel et~al.}(2018)\citenamefont{Raithel, \"Ozel, and
  Psaltis}}]{Raithel:2018ncd}
\bibinfo{author}{\bibfnamefont{C.}~\bibnamefont{Raithel}},
  \bibinfo{author}{\bibfnamefont{F.}~\bibnamefont{\"Ozel}}, \bibnamefont{and}
  \bibinfo{author}{\bibfnamefont{D.}~\bibnamefont{Psaltis}},
  \bibinfo{journal}{Astrophys. J. Lett.} \textbf{\bibinfo{volume}{857}},
  \bibinfo{pages}{L23} (\bibinfo{year}{2018}), \eprint{1803.07687}.

\bibitem[{\citenamefont{De et~al.}(2018)\citenamefont{De, Finstad, Lattimer,
  Brown, Berger, and Biwer}}]{De:2018uhw}
\bibinfo{author}{\bibfnamefont{S.}~\bibnamefont{De}},
  \bibinfo{author}{\bibfnamefont{D.}~\bibnamefont{Finstad}},
  \bibinfo{author}{\bibfnamefont{J.~M.} \bibnamefont{Lattimer}},
  \bibinfo{author}{\bibfnamefont{D.~A.} \bibnamefont{Brown}},
  \bibinfo{author}{\bibfnamefont{E.}~\bibnamefont{Berger}}, \bibnamefont{and}
  \bibinfo{author}{\bibfnamefont{C.~M.} \bibnamefont{Biwer}},
  \bibinfo{journal}{Phys. Rev. Lett.} \textbf{\bibinfo{volume}{121}},
  \bibinfo{pages}{091102} (\bibinfo{year}{2018}), \bibinfo{note}{[Erratum:
  Phys.Rev.Lett. 121, 259902 (2018)]}, \eprint{1804.08583}.

\bibitem[{\citenamefont{Chatziioannou et~al.}(2018)\citenamefont{Chatziioannou,
  Haster, and Zimmerman}}]{Chatziioannou:2018vzf}
\bibinfo{author}{\bibfnamefont{K.}~\bibnamefont{Chatziioannou}},
  \bibinfo{author}{\bibfnamefont{C.-J.} \bibnamefont{Haster}},
  \bibnamefont{and}
  \bibinfo{author}{\bibfnamefont{A.}~\bibnamefont{Zimmerman}},
  \bibinfo{journal}{Phys. Rev. D} \textbf{\bibinfo{volume}{97}},
  \bibinfo{pages}{104036} (\bibinfo{year}{2018}), \eprint{1804.03221}.

\bibitem[{\citenamefont{Carson et~al.}(2019{\natexlab{a}})\citenamefont{Carson,
  Steiner, and Yagi}}]{Carson:2018xri}
\bibinfo{author}{\bibfnamefont{Z.}~\bibnamefont{Carson}},
  \bibinfo{author}{\bibfnamefont{A.~W.} \bibnamefont{Steiner}},
  \bibnamefont{and} \bibinfo{author}{\bibfnamefont{K.}~\bibnamefont{Yagi}},
  \bibinfo{journal}{Phys. Rev. D} \textbf{\bibinfo{volume}{99}},
  \bibinfo{pages}{043010} (\bibinfo{year}{2019}{\natexlab{a}}),
  \eprint{1812.08910}.

\bibitem[{\citenamefont{Abbott et~al.}(2019)}]{LIGOScientific:2018hze}
\bibinfo{author}{\bibfnamefont{B.~P.} \bibnamefont{Abbott}}
  \bibnamefont{et~al.} (\bibinfo{collaboration}{LIGO Scientific, Virgo}),
  \bibinfo{journal}{Phys. Rev. X} \textbf{\bibinfo{volume}{9}},
  \bibinfo{pages}{011001} (\bibinfo{year}{2019}), \eprint{1805.11579}.

\bibitem[{\citenamefont{Carson et~al.}(2019{\natexlab{b}})\citenamefont{Carson,
  Chatziioannou, Haster, Yagi, and Yunes}}]{Carson:2019rjx}
\bibinfo{author}{\bibfnamefont{Z.}~\bibnamefont{Carson}},
  \bibinfo{author}{\bibfnamefont{K.}~\bibnamefont{Chatziioannou}},
  \bibinfo{author}{\bibfnamefont{C.-J.} \bibnamefont{Haster}},
  \bibinfo{author}{\bibfnamefont{K.}~\bibnamefont{Yagi}}, \bibnamefont{and}
  \bibinfo{author}{\bibfnamefont{N.}~\bibnamefont{Yunes}},
  \bibinfo{journal}{Phys. Rev. D} \textbf{\bibinfo{volume}{99}},
  \bibinfo{pages}{083016} (\bibinfo{year}{2019}{\natexlab{b}}),
  \eprint{1903.03909}.

\bibitem[{\citenamefont{Yagi and Yunes}(2016)}]{Yagi:2015pkc}
\bibinfo{author}{\bibfnamefont{K.}~\bibnamefont{Yagi}} \bibnamefont{and}
  \bibinfo{author}{\bibfnamefont{N.}~\bibnamefont{Yunes}},
  \bibinfo{journal}{Class. Quant. Grav.} \textbf{\bibinfo{volume}{33}},
  \bibinfo{pages}{13LT01} (\bibinfo{year}{2016}), \eprint{1512.02639}.

\bibitem[{\citenamefont{Yagi and Yunes}(2017)}]{Yagi:2016bkt}
\bibinfo{author}{\bibfnamefont{K.}~\bibnamefont{Yagi}} \bibnamefont{and}
  \bibinfo{author}{\bibfnamefont{N.}~\bibnamefont{Yunes}},
  \bibinfo{journal}{Phys. Rept.} \textbf{\bibinfo{volume}{681}},
  \bibinfo{pages}{1} (\bibinfo{year}{2017}), \eprint{1608.02582}.

\bibitem[{\citenamefont{Zhao and Lattimer}(2018)}]{Zhao:2018nyf}
\bibinfo{author}{\bibfnamefont{T.}~\bibnamefont{Zhao}} \bibnamefont{and}
  \bibinfo{author}{\bibfnamefont{J.~M.} \bibnamefont{Lattimer}},
  \bibinfo{journal}{Phys. Rev. D} \textbf{\bibinfo{volume}{98}},
  \bibinfo{pages}{063020} (\bibinfo{year}{2018}), \eprint{1808.02858}.

\bibitem[{\citenamefont{Pratten et~al.}(2022)\citenamefont{Pratten, Schmidt,
  and Williams}}]{Pratten:2021pro}
\bibinfo{author}{\bibfnamefont{G.}~\bibnamefont{Pratten}},
  \bibinfo{author}{\bibfnamefont{P.}~\bibnamefont{Schmidt}}, \bibnamefont{and}
  \bibinfo{author}{\bibfnamefont{N.}~\bibnamefont{Williams}},
  \bibinfo{journal}{Phys. Rev. Lett.} \textbf{\bibinfo{volume}{129}},
  \bibinfo{pages}{081102} (\bibinfo{year}{2022}), \eprint{2109.07566}.

\bibitem[{\citenamefont{Gamba and Bernuzzi}(2022)}]{Gamba:2022mgx}
\bibinfo{author}{\bibfnamefont{R.}~\bibnamefont{Gamba}} \bibnamefont{and}
  \bibinfo{author}{\bibfnamefont{S.}~\bibnamefont{Bernuzzi}}
  (\bibinfo{year}{2022}), \eprint{2207.13106}.

\bibitem[{\citenamefont{Paschalidis et~al.}(2018)\citenamefont{Paschalidis,
  Yagi, Alvarez-Castillo, Blaschke, and Sedrakian}}]{Paschalidis:2017qmb}
\bibinfo{author}{\bibfnamefont{V.}~\bibnamefont{Paschalidis}},
  \bibinfo{author}{\bibfnamefont{K.}~\bibnamefont{Yagi}},
  \bibinfo{author}{\bibfnamefont{D.}~\bibnamefont{Alvarez-Castillo}},
  \bibinfo{author}{\bibfnamefont{D.~B.} \bibnamefont{Blaschke}},
  \bibnamefont{and}
  \bibinfo{author}{\bibfnamefont{A.}~\bibnamefont{Sedrakian}},
  \bibinfo{journal}{Phys. Rev. D} \textbf{\bibinfo{volume}{97}},
  \bibinfo{pages}{084038} (\bibinfo{year}{2018}), \eprint{1712.00451}.

\bibitem[{\citenamefont{Christian et~al.}(2019)\citenamefont{Christian, Zacchi,
  and Schaffner-Bielich}}]{Christian:2018jyd}
\bibinfo{author}{\bibfnamefont{J.-E.} \bibnamefont{Christian}},
  \bibinfo{author}{\bibfnamefont{A.}~\bibnamefont{Zacchi}}, \bibnamefont{and}
  \bibinfo{author}{\bibfnamefont{J.}~\bibnamefont{Schaffner-Bielich}},
  \bibinfo{journal}{Phys. Rev. D} \textbf{\bibinfo{volume}{99}},
  \bibinfo{pages}{023009} (\bibinfo{year}{2019}), \eprint{1809.03333}.

\bibitem[{\citenamefont{Dexheimer et~al.}(2019)\citenamefont{Dexheimer,
  de~Oliveira~Gomes, Schramm, and Pais}}]{Dexheimer:2018dhb}
\bibinfo{author}{\bibfnamefont{V.}~\bibnamefont{Dexheimer}},
  \bibinfo{author}{\bibfnamefont{R.}~\bibnamefont{de~Oliveira~Gomes}},
  \bibinfo{author}{\bibfnamefont{S.}~\bibnamefont{Schramm}}, \bibnamefont{and}
  \bibinfo{author}{\bibfnamefont{H.}~\bibnamefont{Pais}}, \bibinfo{journal}{J.
  Phys. G} \textbf{\bibinfo{volume}{46}}, \bibinfo{pages}{034002}
  (\bibinfo{year}{2019}), \eprint{1810.06109}.

\bibitem[{\citenamefont{Han and Steiner}(2019)}]{Han:2018mtj}
\bibinfo{author}{\bibfnamefont{S.}~\bibnamefont{Han}} \bibnamefont{and}
  \bibinfo{author}{\bibfnamefont{A.~W.} \bibnamefont{Steiner}},
  \bibinfo{journal}{Phys. Rev. D} \textbf{\bibinfo{volume}{99}},
  \bibinfo{pages}{083014} (\bibinfo{year}{2019}), \eprint{1810.10967}.

\bibitem[{\citenamefont{Sieniawska et~al.}(2019)\citenamefont{Sieniawska,
  Turczanski, Bejger, and Zdunik}}]{Sieniawska:2018zzj}
\bibinfo{author}{\bibfnamefont{M.}~\bibnamefont{Sieniawska}},
  \bibinfo{author}{\bibfnamefont{W.}~\bibnamefont{Turczanski}},
  \bibinfo{author}{\bibfnamefont{M.}~\bibnamefont{Bejger}}, \bibnamefont{and}
  \bibinfo{author}{\bibfnamefont{J.~L.} \bibnamefont{Zdunik}},
  \bibinfo{journal}{Astron. Astrophys.} \textbf{\bibinfo{volume}{622}},
  \bibinfo{pages}{A174} (\bibinfo{year}{2019}), \eprint{1807.11581}.

\bibitem[{\citenamefont{Chatziioannou and Han}(2020)}]{Chatziioannou:2019yko}
\bibinfo{author}{\bibfnamefont{K.}~\bibnamefont{Chatziioannou}}
  \bibnamefont{and} \bibinfo{author}{\bibfnamefont{S.}~\bibnamefont{Han}},
  \bibinfo{journal}{Phys. Rev. D} \textbf{\bibinfo{volume}{101}},
  \bibinfo{pages}{044019} (\bibinfo{year}{2020}), \eprint{1911.07091}.

\bibitem[{\citenamefont{Tan et~al.}(2022)\citenamefont{Tan, Dexheimer,
  Noronha-Hostler, and Yunes}}]{Tan:2021nat}
\bibinfo{author}{\bibfnamefont{H.}~\bibnamefont{Tan}},
  \bibinfo{author}{\bibfnamefont{V.}~\bibnamefont{Dexheimer}},
  \bibinfo{author}{\bibfnamefont{J.}~\bibnamefont{Noronha-Hostler}},
  \bibnamefont{and} \bibinfo{author}{\bibfnamefont{N.}~\bibnamefont{Yunes}},
  \bibinfo{journal}{Phys. Rev. Lett.} \textbf{\bibinfo{volume}{128}},
  \bibinfo{pages}{161101} (\bibinfo{year}{2022}), \eprint{2111.10260}.

\bibitem[{\citenamefont{Bogdanov et~al.}(2022)}]{Bogdanov:2022faf}
\bibinfo{author}{\bibfnamefont{S.}~\bibnamefont{Bogdanov}}
  \bibnamefont{et~al.}, in \emph{\bibinfo{booktitle}{{2022 Snowmass Summer
  Study}}} (\bibinfo{year}{2022}), \eprint{2209.07412}.

\bibitem[{\citenamefont{Gezerlis et~al.}(2013)\citenamefont{Gezerlis, Tews,
  Epelbaum, Gandolfi, Hebeler, Nogga, and Schwenk}}]{Gezerlis:2013ipa}
\bibinfo{author}{\bibfnamefont{A.}~\bibnamefont{Gezerlis}},
  \bibinfo{author}{\bibfnamefont{I.}~\bibnamefont{Tews}},
  \bibinfo{author}{\bibfnamefont{E.}~\bibnamefont{Epelbaum}},
  \bibinfo{author}{\bibfnamefont{S.}~\bibnamefont{Gandolfi}},
  \bibinfo{author}{\bibfnamefont{K.}~\bibnamefont{Hebeler}},
  \bibinfo{author}{\bibfnamefont{A.}~\bibnamefont{Nogga}}, \bibnamefont{and}
  \bibinfo{author}{\bibfnamefont{A.}~\bibnamefont{Schwenk}},
  \bibinfo{journal}{Phys. Rev. Lett.} \textbf{\bibinfo{volume}{111}},
  \bibinfo{pages}{032501} (\bibinfo{year}{2013}), \eprint{1303.6243}.

\bibitem[{\citenamefont{Lynn et~al.}(2016)\citenamefont{Lynn, Tews, Carlson,
  Gandolfi, Gezerlis, Schmidt, and Schwenk}}]{Lynn:2015jua}
\bibinfo{author}{\bibfnamefont{J.~E.} \bibnamefont{Lynn}},
  \bibinfo{author}{\bibfnamefont{I.}~\bibnamefont{Tews}},
  \bibinfo{author}{\bibfnamefont{J.}~\bibnamefont{Carlson}},
  \bibinfo{author}{\bibfnamefont{S.}~\bibnamefont{Gandolfi}},
  \bibinfo{author}{\bibfnamefont{A.}~\bibnamefont{Gezerlis}},
  \bibinfo{author}{\bibfnamefont{K.~E.} \bibnamefont{Schmidt}},
  \bibnamefont{and} \bibinfo{author}{\bibfnamefont{A.}~\bibnamefont{Schwenk}},
  \bibinfo{journal}{Phys. Rev. Lett.} \textbf{\bibinfo{volume}{116}},
  \bibinfo{pages}{062501} (\bibinfo{year}{2016}), \eprint{1509.03470}.

\bibitem[{\citenamefont{Tews et~al.}(2016)\citenamefont{Tews, Gandolfi,
  Gezerlis, and Schwenk}}]{Tews:2015ufa}
\bibinfo{author}{\bibfnamefont{I.}~\bibnamefont{Tews}},
  \bibinfo{author}{\bibfnamefont{S.}~\bibnamefont{Gandolfi}},
  \bibinfo{author}{\bibfnamefont{A.}~\bibnamefont{Gezerlis}}, \bibnamefont{and}
  \bibinfo{author}{\bibfnamefont{A.}~\bibnamefont{Schwenk}},
  \bibinfo{journal}{Phys. Rev. C} \textbf{\bibinfo{volume}{93}},
  \bibinfo{pages}{024305} (\bibinfo{year}{2016}), \eprint{1507.05561}.

\bibitem[{\citenamefont{Drischler et~al.}(2019)\citenamefont{Drischler,
  Hebeler, and Schwenk}}]{Drischler:2017wtt}
\bibinfo{author}{\bibfnamefont{C.}~\bibnamefont{Drischler}},
  \bibinfo{author}{\bibfnamefont{K.}~\bibnamefont{Hebeler}}, \bibnamefont{and}
  \bibinfo{author}{\bibfnamefont{A.}~\bibnamefont{Schwenk}},
  \bibinfo{journal}{Phys. Rev. Lett.} \textbf{\bibinfo{volume}{122}},
  \bibinfo{pages}{042501} (\bibinfo{year}{2019}), \eprint{1710.08220}.

\bibitem[{\citenamefont{Raithel et~al.}(2017)\citenamefont{Raithel, \"Ozel, and
  Psaltis}}]{Raithel:2017ity}
\bibinfo{author}{\bibfnamefont{C.~A.} \bibnamefont{Raithel}},
  \bibinfo{author}{\bibfnamefont{F.}~\bibnamefont{\"Ozel}}, \bibnamefont{and}
  \bibinfo{author}{\bibfnamefont{D.}~\bibnamefont{Psaltis}},
  \bibinfo{journal}{Astrophys. J.} \textbf{\bibinfo{volume}{844}},
  \bibinfo{pages}{156} (\bibinfo{year}{2017}), \eprint{1704.00737}.

\bibitem[{\citenamefont{Akmal et~al.}(1998)\citenamefont{Akmal, Pandharipande,
  and Ravenhall}}]{Akmal:1998cf}
\bibinfo{author}{\bibfnamefont{A.}~\bibnamefont{Akmal}},
  \bibinfo{author}{\bibfnamefont{V.~R.} \bibnamefont{Pandharipande}},
  \bibnamefont{and} \bibinfo{author}{\bibfnamefont{D.~G.}
  \bibnamefont{Ravenhall}}, \bibinfo{journal}{Phys. Rev. C}
  \textbf{\bibinfo{volume}{58}}, \bibinfo{pages}{1804} (\bibinfo{year}{1998}),
  \eprint{nucl-th/9804027}.

\bibitem[{\citenamefont{Margalit and Metzger}(2017)}]{Margalit:2017dij}
\bibinfo{author}{\bibfnamefont{B.}~\bibnamefont{Margalit}} \bibnamefont{and}
  \bibinfo{author}{\bibfnamefont{B.~D.} \bibnamefont{Metzger}},
  \bibinfo{journal}{Astrophys. J. Lett.} \textbf{\bibinfo{volume}{850}},
  \bibinfo{pages}{L19} (\bibinfo{year}{2017}), \eprint{1710.05938}.

\bibitem[{\citenamefont{Rezzolla et~al.}(2018)\citenamefont{Rezzolla, Most, and
  Weih}}]{Rezzolla:2017aly}
\bibinfo{author}{\bibfnamefont{L.}~\bibnamefont{Rezzolla}},
  \bibinfo{author}{\bibfnamefont{E.~R.} \bibnamefont{Most}}, \bibnamefont{and}
  \bibinfo{author}{\bibfnamefont{L.~R.} \bibnamefont{Weih}},
  \bibinfo{journal}{Astrophys. J. Lett.} \textbf{\bibinfo{volume}{852}},
  \bibinfo{pages}{L25} (\bibinfo{year}{2018}), \eprint{1711.00314}.

\bibitem[{\citenamefont{Ruiz et~al.}(2018)\citenamefont{Ruiz, Shapiro, and
  Tsokaros}}]{Ruiz:2017due}
\bibinfo{author}{\bibfnamefont{M.}~\bibnamefont{Ruiz}},
  \bibinfo{author}{\bibfnamefont{S.~L.} \bibnamefont{Shapiro}},
  \bibnamefont{and} \bibinfo{author}{\bibfnamefont{A.}~\bibnamefont{Tsokaros}},
  \bibinfo{journal}{Phys. Rev. D} \textbf{\bibinfo{volume}{97}},
  \bibinfo{pages}{021501} (\bibinfo{year}{2018}), \eprint{1711.00473}.

\bibitem[{\citenamefont{Shibata et~al.}(2019)\citenamefont{Shibata, Zhou,
  Kiuchi, and Fujibayashi}}]{Shibata:2019ctb}
\bibinfo{author}{\bibfnamefont{M.}~\bibnamefont{Shibata}},
  \bibinfo{author}{\bibfnamefont{E.}~\bibnamefont{Zhou}},
  \bibinfo{author}{\bibfnamefont{K.}~\bibnamefont{Kiuchi}}, \bibnamefont{and}
  \bibinfo{author}{\bibfnamefont{S.}~\bibnamefont{Fujibayashi}},
  \bibinfo{journal}{Phys. Rev. D} \textbf{\bibinfo{volume}{100}},
  \bibinfo{pages}{023015} (\bibinfo{year}{2019}), \eprint{1905.03656}.

\bibitem[{\citenamefont{Nathanail et~al.}(2021)\citenamefont{Nathanail, Most,
  and Rezzolla}}]{Nathanail:2021tay}
\bibinfo{author}{\bibfnamefont{A.}~\bibnamefont{Nathanail}},
  \bibinfo{author}{\bibfnamefont{E.~R.} \bibnamefont{Most}}, \bibnamefont{and}
  \bibinfo{author}{\bibfnamefont{L.}~\bibnamefont{Rezzolla}},
  \bibinfo{journal}{Astrophys. J. Lett.} \textbf{\bibinfo{volume}{908}},
  \bibinfo{pages}{L28} (\bibinfo{year}{2021}), \eprint{2101.01735}.

\bibitem[{\citenamefont{\"Ozel and Freire}(2016)}]{Ozel:2016oaf}
\bibinfo{author}{\bibfnamefont{F.}~\bibnamefont{\"Ozel}} \bibnamefont{and}
  \bibinfo{author}{\bibfnamefont{P.}~\bibnamefont{Freire}},
  \bibinfo{journal}{Ann. Rev. Astron. Astrophys.}
  \textbf{\bibinfo{volume}{54}}, \bibinfo{pages}{401} (\bibinfo{year}{2016}),
  \eprint{1603.02698}.

\bibitem[{\citenamefont{Miller et~al.}(2019)}]{Miller:2019cac}
\bibinfo{author}{\bibfnamefont{M.~C.} \bibnamefont{Miller}}
  \bibnamefont{et~al.}, \bibinfo{journal}{Astrophys. J. Lett.}
  \textbf{\bibinfo{volume}{887}}, \bibinfo{pages}{L24} (\bibinfo{year}{2019}),
  \eprint{1912.05705}.

\bibitem[{\citenamefont{Riley et~al.}(2019)}]{Riley:2019yda}
\bibinfo{author}{\bibfnamefont{T.~E.} \bibnamefont{Riley}}
  \bibnamefont{et~al.}, \bibinfo{journal}{Astrophys. J. Lett.}
  \textbf{\bibinfo{volume}{887}}, \bibinfo{pages}{L21} (\bibinfo{year}{2019}),
  \eprint{1912.05702}.

\bibitem[{\citenamefont{Raaijmakers et~al.}(2019)}]{Raaijmakers:2019qny}
\bibinfo{author}{\bibfnamefont{G.}~\bibnamefont{Raaijmakers}}
  \bibnamefont{et~al.}, \bibinfo{journal}{Astrophys. J. Lett.}
  \textbf{\bibinfo{volume}{887}}, \bibinfo{pages}{L22} (\bibinfo{year}{2019}),
  \eprint{1912.05703}.

\bibitem[{\citenamefont{Raaijmakers et~al.}(2020)}]{Raaijmakers:2019dks}
\bibinfo{author}{\bibfnamefont{G.}~\bibnamefont{Raaijmakers}}
  \bibnamefont{et~al.}, \bibinfo{journal}{Astrophys. J. Lett.}
  \textbf{\bibinfo{volume}{893}}, \bibinfo{pages}{L21} (\bibinfo{year}{2020}),
  \eprint{1912.11031}.

\bibitem[{\citenamefont{Miller et~al.}(2021)}]{Miller:2021qha}
\bibinfo{author}{\bibfnamefont{M.~C.} \bibnamefont{Miller}}
  \bibnamefont{et~al.}, \bibinfo{journal}{Astrophys. J. Lett.}
  \textbf{\bibinfo{volume}{918}}, \bibinfo{pages}{L28} (\bibinfo{year}{2021}),
  \eprint{2105.06979}.

\bibitem[{\citenamefont{Riley et~al.}(2021)}]{Riley:2021pdl}
\bibinfo{author}{\bibfnamefont{T.~E.} \bibnamefont{Riley}}
  \bibnamefont{et~al.}, \bibinfo{journal}{Astrophys. J. Lett.}
  \textbf{\bibinfo{volume}{918}}, \bibinfo{pages}{L27} (\bibinfo{year}{2021}),
  \eprint{2105.06980}.

\bibitem[{\citenamefont{Raaijmakers et~al.}(2021)\citenamefont{Raaijmakers,
  Greif, Hebeler, Hinderer, Nissanke, Schwenk, Riley, Watts, Lattimer, and
  Ho}}]{Raaijmakers:2021uju}
\bibinfo{author}{\bibfnamefont{G.}~\bibnamefont{Raaijmakers}},
  \bibinfo{author}{\bibfnamefont{S.~K.} \bibnamefont{Greif}},
  \bibinfo{author}{\bibfnamefont{K.}~\bibnamefont{Hebeler}},
  \bibinfo{author}{\bibfnamefont{T.}~\bibnamefont{Hinderer}},
  \bibinfo{author}{\bibfnamefont{S.}~\bibnamefont{Nissanke}},
  \bibinfo{author}{\bibfnamefont{A.}~\bibnamefont{Schwenk}},
  \bibinfo{author}{\bibfnamefont{T.~E.} \bibnamefont{Riley}},
  \bibinfo{author}{\bibfnamefont{A.~L.} \bibnamefont{Watts}},
  \bibinfo{author}{\bibfnamefont{J.~M.} \bibnamefont{Lattimer}},
  \bibnamefont{and} \bibinfo{author}{\bibfnamefont{W.~C.~G.} \bibnamefont{Ho}},
  \bibinfo{journal}{Astrophys. J. Lett.} \textbf{\bibinfo{volume}{918}},
  \bibinfo{pages}{L29} (\bibinfo{year}{2021}), \eprint{2105.06981}.

\bibitem[{\citenamefont{Chatziioannou}(2020)}]{Chatziioannou:2020pqz}
\bibinfo{author}{\bibfnamefont{K.}~\bibnamefont{Chatziioannou}},
  \bibinfo{journal}{Gen. Rel. Grav.} \textbf{\bibinfo{volume}{52}},
  \bibinfo{pages}{109} (\bibinfo{year}{2020}), \eprint{2006.03168}.

\bibitem[{\citenamefont{Martinez et~al.}(2015)\citenamefont{Martinez, Stovall,
  Freire, Deneva, Jenet, McLaughlin, Bagchi, Bates, and
  Ridolfi}}]{Martinez:2015mya}
\bibinfo{author}{\bibfnamefont{J.~G.} \bibnamefont{Martinez}},
  \bibinfo{author}{\bibfnamefont{K.}~\bibnamefont{Stovall}},
  \bibinfo{author}{\bibfnamefont{P.~C.~C.} \bibnamefont{Freire}},
  \bibinfo{author}{\bibfnamefont{J.~S.} \bibnamefont{Deneva}},
  \bibinfo{author}{\bibfnamefont{F.~A.} \bibnamefont{Jenet}},
  \bibinfo{author}{\bibfnamefont{M.~A.} \bibnamefont{McLaughlin}},
  \bibinfo{author}{\bibfnamefont{M.}~\bibnamefont{Bagchi}},
  \bibinfo{author}{\bibfnamefont{S.~D.} \bibnamefont{Bates}}, \bibnamefont{and}
  \bibinfo{author}{\bibfnamefont{A.}~\bibnamefont{Ridolfi}},
  \bibinfo{journal}{Astrophys. J.} \textbf{\bibinfo{volume}{812}},
  \bibinfo{pages}{143} (\bibinfo{year}{2015}), \eprint{1509.08805}.

\bibitem[{\citenamefont{Cromartie et~al.}(2019)}]{NANOGrav:2019jur}
\bibinfo{author}{\bibfnamefont{H.~T.} \bibnamefont{Cromartie}}
  \bibnamefont{et~al.} (\bibinfo{collaboration}{NANOGrav}),
  \bibinfo{journal}{Nature Astron.} \textbf{\bibinfo{volume}{4}},
  \bibinfo{pages}{72} (\bibinfo{year}{2019}), \eprint{1904.06759}.

\bibitem[{\citenamefont{Fonseca et~al.}(2021)}]{Fonseca:2021wxt}
\bibinfo{author}{\bibfnamefont{E.}~\bibnamefont{Fonseca}} \bibnamefont{et~al.},
  \bibinfo{journal}{Astrophys. J. Lett.} \textbf{\bibinfo{volume}{915}},
  \bibinfo{pages}{L12} (\bibinfo{year}{2021}), \eprint{2104.00880}.

\bibitem[{\citenamefont{Abbott et~al.}(2017)}]{LIGOScientific:2017vwq}
\bibinfo{author}{\bibfnamefont{B.~P.} \bibnamefont{Abbott}}
  \bibnamefont{et~al.} (\bibinfo{collaboration}{LIGO Scientific, Virgo}),
  \bibinfo{journal}{Phys. Rev. Lett.} \textbf{\bibinfo{volume}{119}},
  \bibinfo{pages}{161101} (\bibinfo{year}{2017}), \eprint{1710.05832}.

\bibitem[{\citenamefont{Jeffreys}(1961)}]{Jeffreys61}
\bibinfo{author}{\bibfnamefont{H.}~\bibnamefont{Jeffreys}},
  \emph{\bibinfo{title}{Theory of Probability}} (\bibinfo{publisher}{Oxford},
  \bibinfo{address}{Oxford, England}, \bibinfo{year}{1961}),
  \bibinfo{edition}{3rd} ed.

\bibitem[{\citenamefont{Gamba et~al.}(2020)\citenamefont{Gamba, Read, and
  Wade}}]{Gamba:2019kwu}
\bibinfo{author}{\bibfnamefont{R.}~\bibnamefont{Gamba}},
  \bibinfo{author}{\bibfnamefont{J.~S.} \bibnamefont{Read}}, \bibnamefont{and}
  \bibinfo{author}{\bibfnamefont{L.~E.} \bibnamefont{Wade}},
  \bibinfo{journal}{Class. Quant. Grav.} \textbf{\bibinfo{volume}{37}},
  \bibinfo{pages}{025008} (\bibinfo{year}{2020}), \eprint{1902.04616}.

\bibitem[{\citenamefont{Reitze et~al.}(2019)}]{Reitze:2019iox}
\bibinfo{author}{\bibfnamefont{D.}~\bibnamefont{Reitze}} \bibnamefont{et~al.},
  \bibinfo{journal}{Bull. Am. Astron. Soc.} \textbf{\bibinfo{volume}{51}},
  \bibinfo{pages}{035} (\bibinfo{year}{2019}), \eprint{1907.04833}.

\bibitem[{\citenamefont{Kojo et~al.}(2015)\citenamefont{Kojo, Powell, Song, and
  Baym}}]{Kojo:2014rca}
\bibinfo{author}{\bibfnamefont{T.}~\bibnamefont{Kojo}},
  \bibinfo{author}{\bibfnamefont{P.~D.} \bibnamefont{Powell}},
  \bibinfo{author}{\bibfnamefont{Y.}~\bibnamefont{Song}}, \bibnamefont{and}
  \bibinfo{author}{\bibfnamefont{G.}~\bibnamefont{Baym}},
  \bibinfo{journal}{Phys. Rev. D} \textbf{\bibinfo{volume}{91}},
  \bibinfo{pages}{045003} (\bibinfo{year}{2015}), \eprint{1412.1108}.

\bibitem[{\citenamefont{Baym et~al.}(2018)\citenamefont{Baym, Hatsuda, Kojo,
  Powell, Song, and Takatsuka}}]{Baym:2017whm}
\bibinfo{author}{\bibfnamefont{G.}~\bibnamefont{Baym}},
  \bibinfo{author}{\bibfnamefont{T.}~\bibnamefont{Hatsuda}},
  \bibinfo{author}{\bibfnamefont{T.}~\bibnamefont{Kojo}},
  \bibinfo{author}{\bibfnamefont{P.~D.} \bibnamefont{Powell}},
  \bibinfo{author}{\bibfnamefont{Y.}~\bibnamefont{Song}}, \bibnamefont{and}
  \bibinfo{author}{\bibfnamefont{T.}~\bibnamefont{Takatsuka}},
  \bibinfo{journal}{Rept. Prog. Phys.} \textbf{\bibinfo{volume}{81}},
  \bibinfo{pages}{056902} (\bibinfo{year}{2018}), \eprint{1707.04966}.

\bibitem[{\citenamefont{Blaschke and Chamel}(2018)}]{Blaschke:2018mqw}
\bibinfo{author}{\bibfnamefont{D.}~\bibnamefont{Blaschke}} \bibnamefont{and}
  \bibinfo{author}{\bibfnamefont{N.}~\bibnamefont{Chamel}},
  \bibinfo{journal}{Astrophys. Space Sci. Libr.}
  \textbf{\bibinfo{volume}{457}}, \bibinfo{pages}{337} (\bibinfo{year}{2018}),
  \eprint{1803.01836}.

\bibitem[{\citenamefont{Raithel and
  Most}(2022{\natexlab{a}})}]{Raithel:2022aee}
\bibinfo{author}{\bibfnamefont{C.~A.} \bibnamefont{Raithel}} \bibnamefont{and}
  \bibinfo{author}{\bibfnamefont{E.~R.} \bibnamefont{Most}}
  (\bibinfo{year}{2022}{\natexlab{a}}), \eprint{2208.04295}.

\bibitem[{\citenamefont{Wijngaarden et~al.}(2022)\citenamefont{Wijngaarden,
  Chatziioannou, Bauswein, Clark, and Cornish}}]{Wijngaarden:2022sah}
\bibinfo{author}{\bibfnamefont{M.}~\bibnamefont{Wijngaarden}},
  \bibinfo{author}{\bibfnamefont{K.}~\bibnamefont{Chatziioannou}},
  \bibinfo{author}{\bibfnamefont{A.}~\bibnamefont{Bauswein}},
  \bibinfo{author}{\bibfnamefont{J.~A.} \bibnamefont{Clark}}, \bibnamefont{and}
  \bibinfo{author}{\bibfnamefont{N.~J.} \bibnamefont{Cornish}},
  \bibinfo{journal}{Phys. Rev. D} \textbf{\bibinfo{volume}{105}},
  \bibinfo{pages}{104019} (\bibinfo{year}{2022}), \eprint{2202.09382}.

\bibitem[{\citenamefont{Breschi
  et~al.}(2022{\natexlab{a}})\citenamefont{Breschi, Gamba, Borhanian, Carullo,
  and Bernuzzi}}]{Breschi:2022ens}
\bibinfo{author}{\bibfnamefont{M.}~\bibnamefont{Breschi}},
  \bibinfo{author}{\bibfnamefont{R.}~\bibnamefont{Gamba}},
  \bibinfo{author}{\bibfnamefont{S.}~\bibnamefont{Borhanian}},
  \bibinfo{author}{\bibfnamefont{G.}~\bibnamefont{Carullo}}, \bibnamefont{and}
  \bibinfo{author}{\bibfnamefont{S.}~\bibnamefont{Bernuzzi}}
  (\bibinfo{year}{2022}{\natexlab{a}}), \eprint{2205.09979}.

\bibitem[{\citenamefont{Yu et~al.}(2022)\citenamefont{Yu, Martynov, Adhikari,
  and Chen}}]{Yu:2022upc}
\bibinfo{author}{\bibfnamefont{H.}~\bibnamefont{Yu}},
  \bibinfo{author}{\bibfnamefont{D.}~\bibnamefont{Martynov}},
  \bibinfo{author}{\bibfnamefont{R.~X.} \bibnamefont{Adhikari}},
  \bibnamefont{and} \bibinfo{author}{\bibfnamefont{Y.}~\bibnamefont{Chen}},
  \bibinfo{journal}{Phys. Rev. D} \textbf{\bibinfo{volume}{106}},
  \bibinfo{pages}{063017} (\bibinfo{year}{2022}), \eprint{2205.14197}.

\bibitem[{\citenamefont{Bauswein et~al.}(2012)\citenamefont{Bauswein, Janka,
  Hebeler, and Schwenk}}]{Bauswein:2012ya}
\bibinfo{author}{\bibfnamefont{A.}~\bibnamefont{Bauswein}},
  \bibinfo{author}{\bibfnamefont{H.~T.} \bibnamefont{Janka}},
  \bibinfo{author}{\bibfnamefont{K.}~\bibnamefont{Hebeler}}, \bibnamefont{and}
  \bibinfo{author}{\bibfnamefont{A.}~\bibnamefont{Schwenk}},
  \bibinfo{journal}{Phys. Rev. D} \textbf{\bibinfo{volume}{86}},
  \bibinfo{pages}{063001} (\bibinfo{year}{2012}), \eprint{1204.1888}.

\bibitem[{\citenamefont{Takami et~al.}(2015)\citenamefont{Takami, Rezzolla, and
  Baiotti}}]{Takami:2014tva}
\bibinfo{author}{\bibfnamefont{K.}~\bibnamefont{Takami}},
  \bibinfo{author}{\bibfnamefont{L.}~\bibnamefont{Rezzolla}}, \bibnamefont{and}
  \bibinfo{author}{\bibfnamefont{L.}~\bibnamefont{Baiotti}},
  \bibinfo{journal}{Phys. Rev. D} \textbf{\bibinfo{volume}{91}},
  \bibinfo{pages}{064001} (\bibinfo{year}{2015}), \eprint{1412.3240}.

\bibitem[{\citenamefont{Takami et~al.}(2014)\citenamefont{Takami, Rezzolla, and
  Baiotti}}]{Takami:2014zpa}
\bibinfo{author}{\bibfnamefont{K.}~\bibnamefont{Takami}},
  \bibinfo{author}{\bibfnamefont{L.}~\bibnamefont{Rezzolla}}, \bibnamefont{and}
  \bibinfo{author}{\bibfnamefont{L.}~\bibnamefont{Baiotti}},
  \bibinfo{journal}{Phys. Rev. Lett.} \textbf{\bibinfo{volume}{113}},
  \bibinfo{pages}{091104} (\bibinfo{year}{2014}), \eprint{1403.5672}.

\bibitem[{\citenamefont{Baiotti and Rezzolla}(2017)}]{Baiotti:2016qnr}
\bibinfo{author}{\bibfnamefont{L.}~\bibnamefont{Baiotti}} \bibnamefont{and}
  \bibinfo{author}{\bibfnamefont{L.}~\bibnamefont{Rezzolla}},
  \bibinfo{journal}{Rept. Prog. Phys.} \textbf{\bibinfo{volume}{80}},
  \bibinfo{pages}{096901} (\bibinfo{year}{2017}), \eprint{1607.03540}.

\bibitem[{\citenamefont{Paschalidis and
  Stergioulas}(2017)}]{Paschalidis:2016vmz}
\bibinfo{author}{\bibfnamefont{V.}~\bibnamefont{Paschalidis}} \bibnamefont{and}
  \bibinfo{author}{\bibfnamefont{N.}~\bibnamefont{Stergioulas}},
  \bibinfo{journal}{Living Rev. Rel.} \textbf{\bibinfo{volume}{20}},
  \bibinfo{pages}{7} (\bibinfo{year}{2017}), \eprint{1612.03050}.

\bibitem[{\citenamefont{Rezzolla and Takami}(2016)}]{Rezzolla:2016nxn}
\bibinfo{author}{\bibfnamefont{L.}~\bibnamefont{Rezzolla}} \bibnamefont{and}
  \bibinfo{author}{\bibfnamefont{K.}~\bibnamefont{Takami}},
  \bibinfo{journal}{Phys. Rev. D} \textbf{\bibinfo{volume}{93}},
  \bibinfo{pages}{124051} (\bibinfo{year}{2016}), \eprint{1604.00246}.

\bibitem[{\citenamefont{Bauswein and Stergioulas}(2019)}]{Bauswein:2019ybt}
\bibinfo{author}{\bibfnamefont{A.}~\bibnamefont{Bauswein}} \bibnamefont{and}
  \bibinfo{author}{\bibfnamefont{N.}~\bibnamefont{Stergioulas}},
  \bibinfo{journal}{J. Phys. G} \textbf{\bibinfo{volume}{46}},
  \bibinfo{pages}{113002} (\bibinfo{year}{2019}), \eprint{1901.06969}.

\bibitem[{\citenamefont{Bernuzzi}(2020)}]{Bernuzzi:2020tgt}
\bibinfo{author}{\bibfnamefont{S.}~\bibnamefont{Bernuzzi}},
  \bibinfo{journal}{Gen. Rel. Grav.} \textbf{\bibinfo{volume}{52}},
  \bibinfo{pages}{108} (\bibinfo{year}{2020}), \eprint{2004.06419}.

\bibitem[{\citenamefont{Radice et~al.}(2020)\citenamefont{Radice, Bernuzzi, and
  Perego}}]{Radice:2020ddv}
\bibinfo{author}{\bibfnamefont{D.}~\bibnamefont{Radice}},
  \bibinfo{author}{\bibfnamefont{S.}~\bibnamefont{Bernuzzi}}, \bibnamefont{and}
  \bibinfo{author}{\bibfnamefont{A.}~\bibnamefont{Perego}},
  \bibinfo{journal}{Ann. Rev. Nucl. Part. Sci.} \textbf{\bibinfo{volume}{70}},
  \bibinfo{pages}{95} (\bibinfo{year}{2020}), \eprint{2002.03863}.

\bibitem[{\citenamefont{Vretinaris et~al.}(2020)\citenamefont{Vretinaris,
  Stergioulas, and Bauswein}}]{Vretinaris:2019spn}
\bibinfo{author}{\bibfnamefont{S.}~\bibnamefont{Vretinaris}},
  \bibinfo{author}{\bibfnamefont{N.}~\bibnamefont{Stergioulas}},
  \bibnamefont{and} \bibinfo{author}{\bibfnamefont{A.}~\bibnamefont{Bauswein}},
  \bibinfo{journal}{Phys. Rev. D} \textbf{\bibinfo{volume}{101}},
  \bibinfo{pages}{084039} (\bibinfo{year}{2020}), \eprint{1910.10856}.

\bibitem[{\citenamefont{Raithel and
  Most}(2022{\natexlab{b}})}]{Raithel:2022orm}
\bibinfo{author}{\bibfnamefont{C.~A.} \bibnamefont{Raithel}} \bibnamefont{and}
  \bibinfo{author}{\bibfnamefont{E.~R.} \bibnamefont{Most}},
  \bibinfo{journal}{Astrophys. J. Lett.} \textbf{\bibinfo{volume}{933}},
  \bibinfo{pages}{L39} (\bibinfo{year}{2022}{\natexlab{b}}),
  \eprint{2201.03594}.

\bibitem[{\citenamefont{Most et~al.}(2019{\natexlab{a}})\citenamefont{Most,
  Papenfort, Dexheimer, Hanauske, Schramm, St\"ocker, and
  Rezzolla}}]{Most:2018eaw}
\bibinfo{author}{\bibfnamefont{E.~R.} \bibnamefont{Most}},
  \bibinfo{author}{\bibfnamefont{L.~J.} \bibnamefont{Papenfort}},
  \bibinfo{author}{\bibfnamefont{V.}~\bibnamefont{Dexheimer}},
  \bibinfo{author}{\bibfnamefont{M.}~\bibnamefont{Hanauske}},
  \bibinfo{author}{\bibfnamefont{S.}~\bibnamefont{Schramm}},
  \bibinfo{author}{\bibfnamefont{H.}~\bibnamefont{St\"ocker}},
  \bibnamefont{and} \bibinfo{author}{\bibfnamefont{L.}~\bibnamefont{Rezzolla}},
  \bibinfo{journal}{Phys. Rev. Lett.} \textbf{\bibinfo{volume}{122}},
  \bibinfo{pages}{061101} (\bibinfo{year}{2019}{\natexlab{a}}),
  \eprint{1807.03684}.

\bibitem[{\citenamefont{Bauswein et~al.}(2019)\citenamefont{Bauswein, Bastian,
  Blaschke, Chatziioannou, Clark, Fischer, and Oertel}}]{Bauswein:2018bma}
\bibinfo{author}{\bibfnamefont{A.}~\bibnamefont{Bauswein}},
  \bibinfo{author}{\bibfnamefont{N.-U.~F.} \bibnamefont{Bastian}},
  \bibinfo{author}{\bibfnamefont{D.~B.} \bibnamefont{Blaschke}},
  \bibinfo{author}{\bibfnamefont{K.}~\bibnamefont{Chatziioannou}},
  \bibinfo{author}{\bibfnamefont{J.~A.} \bibnamefont{Clark}},
  \bibinfo{author}{\bibfnamefont{T.}~\bibnamefont{Fischer}}, \bibnamefont{and}
  \bibinfo{author}{\bibfnamefont{M.}~\bibnamefont{Oertel}},
  \bibinfo{journal}{Phys. Rev. Lett.} \textbf{\bibinfo{volume}{122}},
  \bibinfo{pages}{061102} (\bibinfo{year}{2019}), \eprint{1809.01116}.

\bibitem[{\citenamefont{Weih et~al.}(2020)\citenamefont{Weih, Hanauske, and
  Rezzolla}}]{Weih:2019xvw}
\bibinfo{author}{\bibfnamefont{L.~R.} \bibnamefont{Weih}},
  \bibinfo{author}{\bibfnamefont{M.}~\bibnamefont{Hanauske}}, \bibnamefont{and}
  \bibinfo{author}{\bibfnamefont{L.}~\bibnamefont{Rezzolla}},
  \bibinfo{journal}{Phys. Rev. Lett.} \textbf{\bibinfo{volume}{124}},
  \bibinfo{pages}{171103} (\bibinfo{year}{2020}), \eprint{1912.09340}.

\bibitem[{\citenamefont{Most et~al.}(2020)\citenamefont{Most, Jens~Papenfort,
  Dexheimer, Hanauske, Stoecker, and Rezzolla}}]{Most:2019onn}
\bibinfo{author}{\bibfnamefont{E.~R.} \bibnamefont{Most}},
  \bibinfo{author}{\bibfnamefont{L.}~\bibnamefont{Jens~Papenfort}},
  \bibinfo{author}{\bibfnamefont{V.}~\bibnamefont{Dexheimer}},
  \bibinfo{author}{\bibfnamefont{M.}~\bibnamefont{Hanauske}},
  \bibinfo{author}{\bibfnamefont{H.}~\bibnamefont{Stoecker}}, \bibnamefont{and}
  \bibinfo{author}{\bibfnamefont{L.}~\bibnamefont{Rezzolla}},
  \bibinfo{journal}{Eur. Phys. J. A} \textbf{\bibinfo{volume}{56}},
  \bibinfo{pages}{59} (\bibinfo{year}{2020}), \eprint{1910.13893}.

\bibitem[{\citenamefont{Prakash et~al.}(2021)\citenamefont{Prakash, Radice,
  Logoteta, Perego, Nedora, Bombaci, Kashyap, Bernuzzi, and
  Endrizzi}}]{Prakash:2021wpz}
\bibinfo{author}{\bibfnamefont{A.}~\bibnamefont{Prakash}},
  \bibinfo{author}{\bibfnamefont{D.}~\bibnamefont{Radice}},
  \bibinfo{author}{\bibfnamefont{D.}~\bibnamefont{Logoteta}},
  \bibinfo{author}{\bibfnamefont{A.}~\bibnamefont{Perego}},
  \bibinfo{author}{\bibfnamefont{V.}~\bibnamefont{Nedora}},
  \bibinfo{author}{\bibfnamefont{I.}~\bibnamefont{Bombaci}},
  \bibinfo{author}{\bibfnamefont{R.}~\bibnamefont{Kashyap}},
  \bibinfo{author}{\bibfnamefont{S.}~\bibnamefont{Bernuzzi}}, \bibnamefont{and}
  \bibinfo{author}{\bibfnamefont{A.}~\bibnamefont{Endrizzi}},
  \bibinfo{journal}{Phys. Rev. D} \textbf{\bibinfo{volume}{104}},
  \bibinfo{pages}{083029} (\bibinfo{year}{2021}), \eprint{2106.07885}.

\bibitem[{\citenamefont{Kedia et~al.}(2022)\citenamefont{Kedia, Kim, Suh, and
  Mathews}}]{Kedia:2022nns}
\bibinfo{author}{\bibfnamefont{A.}~\bibnamefont{Kedia}},
  \bibinfo{author}{\bibfnamefont{H.~I.} \bibnamefont{Kim}},
  \bibinfo{author}{\bibfnamefont{I.-S.} \bibnamefont{Suh}}, \bibnamefont{and}
  \bibinfo{author}{\bibfnamefont{G.~J.} \bibnamefont{Mathews}},
  \bibinfo{journal}{Phys. Rev. D} \textbf{\bibinfo{volume}{106}},
  \bibinfo{pages}{103027} (\bibinfo{year}{2022}), \eprint{2203.05461}.

\bibitem[{\citenamefont{Huang et~al.}(2022)\citenamefont{Huang, Baiotti, Kojo,
  Takami, Sotani, Togashi, Hatsuda, Nagataki, and Fan}}]{Huang:2022mqp}
\bibinfo{author}{\bibfnamefont{Y.-J.} \bibnamefont{Huang}},
  \bibinfo{author}{\bibfnamefont{L.}~\bibnamefont{Baiotti}},
  \bibinfo{author}{\bibfnamefont{T.}~\bibnamefont{Kojo}},
  \bibinfo{author}{\bibfnamefont{K.}~\bibnamefont{Takami}},
  \bibinfo{author}{\bibfnamefont{H.}~\bibnamefont{Sotani}},
  \bibinfo{author}{\bibfnamefont{H.}~\bibnamefont{Togashi}},
  \bibinfo{author}{\bibfnamefont{T.}~\bibnamefont{Hatsuda}},
  \bibinfo{author}{\bibfnamefont{S.}~\bibnamefont{Nagataki}}, \bibnamefont{and}
  \bibinfo{author}{\bibfnamefont{Y.-Z.} \bibnamefont{Fan}},
  \bibinfo{journal}{Phys. Rev. Lett.} \textbf{\bibinfo{volume}{129}},
  \bibinfo{pages}{181101} (\bibinfo{year}{2022}), \eprint{2203.04528}.

\bibitem[{\citenamefont{Raithel et~al.}(2019)\citenamefont{Raithel, Ozel, and
  Psaltis}}]{Raithel:2019gws}
\bibinfo{author}{\bibfnamefont{C.~A.} \bibnamefont{Raithel}},
  \bibinfo{author}{\bibfnamefont{F.}~\bibnamefont{Ozel}}, \bibnamefont{and}
  \bibinfo{author}{\bibfnamefont{D.}~\bibnamefont{Psaltis}},
  \bibinfo{journal}{Astrophys. J.} \textbf{\bibinfo{volume}{875}},
  \bibinfo{pages}{12} (\bibinfo{year}{2019}), \eprint{1902.10735}.

\bibitem[{\citenamefont{Most and Raithel}(2021)}]{Most:2021ktk}
\bibinfo{author}{\bibfnamefont{E.~R.} \bibnamefont{Most}} \bibnamefont{and}
  \bibinfo{author}{\bibfnamefont{C.~A.} \bibnamefont{Raithel}},
  \bibinfo{journal}{Phys. Rev. D} \textbf{\bibinfo{volume}{104}},
  \bibinfo{pages}{124012} (\bibinfo{year}{2021}), \eprint{2107.06804}.

\bibitem[{\citenamefont{Most et~al.}(2019{\natexlab{b}})\citenamefont{Most,
  Papenfort, and Rezzolla}}]{Most:2019kfe}
\bibinfo{author}{\bibfnamefont{E.~R.} \bibnamefont{Most}},
  \bibinfo{author}{\bibfnamefont{L.~J.} \bibnamefont{Papenfort}},
  \bibnamefont{and} \bibinfo{author}{\bibfnamefont{L.}~\bibnamefont{Rezzolla}},
  \bibinfo{journal}{Mon. Not. Roy. Astron. Soc.}
  \textbf{\bibinfo{volume}{490}}, \bibinfo{pages}{3588}
  (\bibinfo{year}{2019}{\natexlab{b}}), \eprint{1907.10328}.

\bibitem[{\citenamefont{Etienne et~al.}(2015)\citenamefont{Etienne,
  Paschalidis, Haas, M\"osta, and Shapiro}}]{Etienne:2015cea}
\bibinfo{author}{\bibfnamefont{Z.~B.} \bibnamefont{Etienne}},
  \bibinfo{author}{\bibfnamefont{V.}~\bibnamefont{Paschalidis}},
  \bibinfo{author}{\bibfnamefont{R.}~\bibnamefont{Haas}},
  \bibinfo{author}{\bibfnamefont{P.}~\bibnamefont{M\"osta}}, \bibnamefont{and}
  \bibinfo{author}{\bibfnamefont{S.~L.} \bibnamefont{Shapiro}},
  \bibinfo{journal}{Class. Quant. Grav.} \textbf{\bibinfo{volume}{32}},
  \bibinfo{pages}{175009} (\bibinfo{year}{2015}), \eprint{1501.07276}.

\bibitem[{\citenamefont{Breschi et~al.}(2019)\citenamefont{Breschi, Bernuzzi,
  Zappa, Agathos, Perego, Radice, and Nagar}}]{Breschi:2019srl}
\bibinfo{author}{\bibfnamefont{M.}~\bibnamefont{Breschi}},
  \bibinfo{author}{\bibfnamefont{S.}~\bibnamefont{Bernuzzi}},
  \bibinfo{author}{\bibfnamefont{F.}~\bibnamefont{Zappa}},
  \bibinfo{author}{\bibfnamefont{M.}~\bibnamefont{Agathos}},
  \bibinfo{author}{\bibfnamefont{A.}~\bibnamefont{Perego}},
  \bibinfo{author}{\bibfnamefont{D.}~\bibnamefont{Radice}}, \bibnamefont{and}
  \bibinfo{author}{\bibfnamefont{A.}~\bibnamefont{Nagar}},
  \bibinfo{journal}{Phys. Rev. D} \textbf{\bibinfo{volume}{100}},
  \bibinfo{pages}{104029} (\bibinfo{year}{2019}), \eprint{1908.11418}.

\bibitem[{\citenamefont{Breschi
  et~al.}(2022{\natexlab{b}})\citenamefont{Breschi, Bernuzzi, Chakravarti,
  Camilletti, Prakash, and Perego}}]{Breschi:2022xnc}
\bibinfo{author}{\bibfnamefont{M.}~\bibnamefont{Breschi}},
  \bibinfo{author}{\bibfnamefont{S.}~\bibnamefont{Bernuzzi}},
  \bibinfo{author}{\bibfnamefont{K.}~\bibnamefont{Chakravarti}},
  \bibinfo{author}{\bibfnamefont{A.}~\bibnamefont{Camilletti}},
  \bibinfo{author}{\bibfnamefont{A.}~\bibnamefont{Prakash}}, \bibnamefont{and}
  \bibinfo{author}{\bibfnamefont{A.}~\bibnamefont{Perego}}
  (\bibinfo{year}{2022}{\natexlab{b}}), \eprint{2205.09112}.

\bibitem[{\citenamefont{Bauswein et~al.}(2010)\citenamefont{Bauswein, Janka,
  and Oechslin}}]{Bauswein:2010dn}
\bibinfo{author}{\bibfnamefont{A.}~\bibnamefont{Bauswein}},
  \bibinfo{author}{\bibfnamefont{H.~T.} \bibnamefont{Janka}}, \bibnamefont{and}
  \bibinfo{author}{\bibfnamefont{R.}~\bibnamefont{Oechslin}},
  \bibinfo{journal}{Phys. Rev. D} \textbf{\bibinfo{volume}{82}},
  \bibinfo{pages}{084043} (\bibinfo{year}{2010}), \eprint{1006.3315}.

\bibitem[{\citenamefont{Figura et~al.}(2020)\citenamefont{Figura, Lu, Burgio,
  Li, and Schulze}}]{Figura:2020fkj}
\bibinfo{author}{\bibfnamefont{A.}~\bibnamefont{Figura}},
  \bibinfo{author}{\bibfnamefont{J.~J.} \bibnamefont{Lu}},
  \bibinfo{author}{\bibfnamefont{G.~F.} \bibnamefont{Burgio}},
  \bibinfo{author}{\bibfnamefont{Z.~H.} \bibnamefont{Li}}, \bibnamefont{and}
  \bibinfo{author}{\bibfnamefont{H.~J.} \bibnamefont{Schulze}},
  \bibinfo{journal}{Phys. Rev. D} \textbf{\bibinfo{volume}{102}},
  \bibinfo{pages}{043006} (\bibinfo{year}{2020}), \eprint{2005.08691}.

\bibitem[{\citenamefont{Raithel et~al.}(2021)\citenamefont{Raithel,
  Paschalidis, and \"Ozel}}]{Raithel:2021hye}
\bibinfo{author}{\bibfnamefont{C.}~\bibnamefont{Raithel}},
  \bibinfo{author}{\bibfnamefont{V.}~\bibnamefont{Paschalidis}},
  \bibnamefont{and} \bibinfo{author}{\bibfnamefont{F.}~\bibnamefont{\"Ozel}},
  \bibinfo{journal}{Phys. Rev. D} \textbf{\bibinfo{volume}{104}},
  \bibinfo{pages}{063016} (\bibinfo{year}{2021}), \eprint{2104.07226}.

\bibitem[{\citenamefont{Alford et~al.}(2018)\citenamefont{Alford, Bovard,
  Hanauske, Rezzolla, and Schwenzer}}]{Alford:2017rxf}
\bibinfo{author}{\bibfnamefont{M.~G.} \bibnamefont{Alford}},
  \bibinfo{author}{\bibfnamefont{L.}~\bibnamefont{Bovard}},
  \bibinfo{author}{\bibfnamefont{M.}~\bibnamefont{Hanauske}},
  \bibinfo{author}{\bibfnamefont{L.}~\bibnamefont{Rezzolla}}, \bibnamefont{and}
  \bibinfo{author}{\bibfnamefont{K.}~\bibnamefont{Schwenzer}},
  \bibinfo{journal}{Phys. Rev. Lett.} \textbf{\bibinfo{volume}{120}},
  \bibinfo{pages}{041101} (\bibinfo{year}{2018}), \eprint{1707.09475}.

\bibitem[{\citenamefont{Most et~al.}(2021)\citenamefont{Most, Harris, Plumberg,
  Alford, Noronha, Noronha-Hostler, Pretorius, Witek, and
  Yunes}}]{Most:2021zvc}
\bibinfo{author}{\bibfnamefont{E.~R.} \bibnamefont{Most}},
  \bibinfo{author}{\bibfnamefont{S.~P.} \bibnamefont{Harris}},
  \bibinfo{author}{\bibfnamefont{C.}~\bibnamefont{Plumberg}},
  \bibinfo{author}{\bibfnamefont{M.~G.} \bibnamefont{Alford}},
  \bibinfo{author}{\bibfnamefont{J.}~\bibnamefont{Noronha}},
  \bibinfo{author}{\bibfnamefont{J.}~\bibnamefont{Noronha-Hostler}},
  \bibinfo{author}{\bibfnamefont{F.}~\bibnamefont{Pretorius}},
  \bibinfo{author}{\bibfnamefont{H.}~\bibnamefont{Witek}}, \bibnamefont{and}
  \bibinfo{author}{\bibfnamefont{N.}~\bibnamefont{Yunes}},
  \bibinfo{journal}{Mon. Not. Roy. Astron. Soc.}
  \textbf{\bibinfo{volume}{509}}, \bibinfo{pages}{1096} (\bibinfo{year}{2021}),
  \eprint{2107.05094}.

\bibitem[{\citenamefont{Radice et~al.}(2022)\citenamefont{Radice, Bernuzzi,
  Perego, and Haas}}]{Radice:2021jtw}
\bibinfo{author}{\bibfnamefont{D.}~\bibnamefont{Radice}},
  \bibinfo{author}{\bibfnamefont{S.}~\bibnamefont{Bernuzzi}},
  \bibinfo{author}{\bibfnamefont{A.}~\bibnamefont{Perego}}, \bibnamefont{and}
  \bibinfo{author}{\bibfnamefont{R.}~\bibnamefont{Haas}},
  \bibinfo{journal}{Mon. Not. Roy. Astron. Soc.}
  \textbf{\bibinfo{volume}{512}}, \bibinfo{pages}{1499} (\bibinfo{year}{2022}),
  \eprint{2111.14858}.

\bibitem[{\citenamefont{Most et~al.}(2022)\citenamefont{Most, Haber, Harris,
  Zhang, Alford, and Noronha}}]{Most:2022yhe}
\bibinfo{author}{\bibfnamefont{E.~R.} \bibnamefont{Most}},
  \bibinfo{author}{\bibfnamefont{A.}~\bibnamefont{Haber}},
  \bibinfo{author}{\bibfnamefont{S.~P.} \bibnamefont{Harris}},
  \bibinfo{author}{\bibfnamefont{Z.}~\bibnamefont{Zhang}},
  \bibinfo{author}{\bibfnamefont{M.~G.} \bibnamefont{Alford}},
  \bibnamefont{and} \bibinfo{author}{\bibfnamefont{J.}~\bibnamefont{Noronha}}
  (\bibinfo{year}{2022}), \eprint{2207.00442}.

\bibitem[{\citenamefont{Punturo et~al.}(2010)}]{Punturo:2010zz}
\bibinfo{author}{\bibfnamefont{M.}~\bibnamefont{Punturo}} \bibnamefont{et~al.},
  \bibinfo{journal}{Class. Quant. Grav.} \textbf{\bibinfo{volume}{27}},
  \bibinfo{pages}{194002} (\bibinfo{year}{2010}).

\bibitem[{\citenamefont{Ackley et~al.}(2020)}]{Ackley:2020atn}
\bibinfo{author}{\bibfnamefont{K.}~\bibnamefont{Ackley}} \bibnamefont{et~al.},
  \bibinfo{journal}{Publ. Astron. Soc. Austral.} \textbf{\bibinfo{volume}{37}},
  \bibinfo{pages}{e047} (\bibinfo{year}{2020}), \eprint{2007.03128}.

\bibitem[{\citenamefont{Drischler et~al.}(2021)\citenamefont{Drischler, Holt,
  and Wellenhofer}}]{Drischler:2021kxf}
\bibinfo{author}{\bibfnamefont{C.}~\bibnamefont{Drischler}},
  \bibinfo{author}{\bibfnamefont{J.~W.} \bibnamefont{Holt}}, \bibnamefont{and}
  \bibinfo{author}{\bibfnamefont{C.}~\bibnamefont{Wellenhofer}},
  \bibinfo{journal}{Ann. Rev. Nucl. Part. Sci.} \textbf{\bibinfo{volume}{71}},
  \bibinfo{pages}{403} (\bibinfo{year}{2021}), \eprint{2101.01709}.

\bibitem[{\citenamefont{Tews et~al.}(2022)}]{Tews:2022yfb}
\bibinfo{author}{\bibfnamefont{I.}~\bibnamefont{Tews}} \bibnamefont{et~al.},
  \bibinfo{journal}{Few Body Syst.} \textbf{\bibinfo{volume}{63}},
  \bibinfo{pages}{67} (\bibinfo{year}{2022}), \eprint{2202.01105}.

\bibitem[{\citenamefont{Towns et~al.}(2014)}]{Towns:2014qtb}
\bibinfo{author}{\bibfnamefont{J.}~\bibnamefont{Towns}} \bibnamefont{et~al.},
  \bibinfo{journal}{Comput. Sci. Eng.} \textbf{\bibinfo{volume}{16}},
  \bibinfo{pages}{62} (\bibinfo{year}{2014}).

\bibitem[{\citenamefont{Read et~al.}(2009)\citenamefont{Read, Lackey, Owen, and
  Friedman}}]{Read:2008iy}
\bibinfo{author}{\bibfnamefont{J.~S.} \bibnamefont{Read}},
  \bibinfo{author}{\bibfnamefont{B.~D.} \bibnamefont{Lackey}},
  \bibinfo{author}{\bibfnamefont{B.~J.} \bibnamefont{Owen}}, \bibnamefont{and}
  \bibinfo{author}{\bibfnamefont{J.~L.} \bibnamefont{Friedman}},
  \bibinfo{journal}{Phys. Rev. D} \textbf{\bibinfo{volume}{79}},
  \bibinfo{pages}{124032} (\bibinfo{year}{2009}), \eprint{0812.2163}.

\bibitem[{\citenamefont{Ozel and Psaltis}(2009)}]{Ozel:2009da}
\bibinfo{author}{\bibfnamefont{F.}~\bibnamefont{Ozel}} \bibnamefont{and}
  \bibinfo{author}{\bibfnamefont{D.}~\bibnamefont{Psaltis}},
  \bibinfo{journal}{Phys. Rev. D} \textbf{\bibinfo{volume}{80}},
  \bibinfo{pages}{103003} (\bibinfo{year}{2009}), \eprint{0905.1959}.

\bibitem[{\citenamefont{Raithel et~al.}(2016)\citenamefont{Raithel, Ozel, and
  Psaltis}}]{Raithel:2016bux}
\bibinfo{author}{\bibfnamefont{C.~A.} \bibnamefont{Raithel}},
  \bibinfo{author}{\bibfnamefont{F.}~\bibnamefont{Ozel}}, \bibnamefont{and}
  \bibinfo{author}{\bibfnamefont{D.}~\bibnamefont{Psaltis}},
  \bibinfo{journal}{Astrophys. J.} \textbf{\bibinfo{volume}{831}},
  \bibinfo{pages}{44} (\bibinfo{year}{2016}), \eprint{1605.03591}.

\bibitem[{\citenamefont{Steiner et~al.}(2013)\citenamefont{Steiner, Hempel, and
  Fischer}}]{Steiner:2012rk}
\bibinfo{author}{\bibfnamefont{A.~W.} \bibnamefont{Steiner}},
  \bibinfo{author}{\bibfnamefont{M.}~\bibnamefont{Hempel}}, \bibnamefont{and}
  \bibinfo{author}{\bibfnamefont{T.}~\bibnamefont{Fischer}},
  \bibinfo{journal}{Astrophys. J.} \textbf{\bibinfo{volume}{774}},
  \bibinfo{pages}{17} (\bibinfo{year}{2013}), \eprint{1207.2184}.

\bibitem[{\citenamefont{Raithel et~al.}(2022)\citenamefont{Raithel, Espino, and
  Paschalidis}}]{Raithel:2022nab}
\bibinfo{author}{\bibfnamefont{C.}~\bibnamefont{Raithel}},
  \bibinfo{author}{\bibfnamefont{P.}~\bibnamefont{Espino}}, \bibnamefont{and}
  \bibinfo{author}{\bibfnamefont{V.}~\bibnamefont{Paschalidis}}
  (\bibinfo{year}{2022}), \eprint{2206.14838}.

\bibitem[{\citenamefont{Schneider et~al.}(2017)\citenamefont{Schneider,
  Roberts, and Ott}}]{Schneider:2017tfi}
\bibinfo{author}{\bibfnamefont{A.~S.} \bibnamefont{Schneider}},
  \bibinfo{author}{\bibfnamefont{L.~F.} \bibnamefont{Roberts}},
  \bibnamefont{and} \bibinfo{author}{\bibfnamefont{C.~D.} \bibnamefont{Ott}},
  \bibinfo{journal}{Phys. Rev. C} \textbf{\bibinfo{volume}{96}},
  \bibinfo{pages}{065802} (\bibinfo{year}{2017}), \eprint{1707.01527}.

\bibitem[{\citenamefont{Duez et~al.}(2005)\citenamefont{Duez, Liu, Shapiro, and
  Stephens}}]{Duez:2005sf}
\bibinfo{author}{\bibfnamefont{M.~D.} \bibnamefont{Duez}},
  \bibinfo{author}{\bibfnamefont{Y.~T.} \bibnamefont{Liu}},
  \bibinfo{author}{\bibfnamefont{S.~L.} \bibnamefont{Shapiro}},
  \bibnamefont{and} \bibinfo{author}{\bibfnamefont{B.~C.}
  \bibnamefont{Stephens}}, \bibinfo{journal}{Phys. Rev. D}
  \textbf{\bibinfo{volume}{72}}, \bibinfo{pages}{024028}
  (\bibinfo{year}{2005}), \eprint{astro-ph/0503420}.

\bibitem[{\citenamefont{Shibata and Sekiguchi}(2005)}]{Shibata:2005gp}
\bibinfo{author}{\bibfnamefont{M.}~\bibnamefont{Shibata}} \bibnamefont{and}
  \bibinfo{author}{\bibfnamefont{Y.-i.} \bibnamefont{Sekiguchi}},
  \bibinfo{journal}{Phys. Rev. D} \textbf{\bibinfo{volume}{72}},
  \bibinfo{pages}{044014} (\bibinfo{year}{2005}), \eprint{astro-ph/0507383}.

\bibitem[{\citenamefont{Hilditch et~al.}(2013)\citenamefont{Hilditch, Bernuzzi,
  Thierfelder, Cao, Tichy, and Bruegmann}}]{Hilditch:2012fp}
\bibinfo{author}{\bibfnamefont{D.}~\bibnamefont{Hilditch}},
  \bibinfo{author}{\bibfnamefont{S.}~\bibnamefont{Bernuzzi}},
  \bibinfo{author}{\bibfnamefont{M.}~\bibnamefont{Thierfelder}},
  \bibinfo{author}{\bibfnamefont{Z.}~\bibnamefont{Cao}},
  \bibinfo{author}{\bibfnamefont{W.}~\bibnamefont{Tichy}}, \bibnamefont{and}
  \bibinfo{author}{\bibfnamefont{B.}~\bibnamefont{Bruegmann}},
  \bibinfo{journal}{Phys. Rev. D} \textbf{\bibinfo{volume}{88}},
  \bibinfo{pages}{084057} (\bibinfo{year}{2013}), \eprint{1212.2901}.

\bibitem[{\citenamefont{Bernuzzi and Hilditch}(2010)}]{Bernuzzi:2009ex}
\bibinfo{author}{\bibfnamefont{S.}~\bibnamefont{Bernuzzi}} \bibnamefont{and}
  \bibinfo{author}{\bibfnamefont{D.}~\bibnamefont{Hilditch}},
  \bibinfo{journal}{Phys. Rev. D} \textbf{\bibinfo{volume}{81}},
  \bibinfo{pages}{084003} (\bibinfo{year}{2010}), \eprint{0912.2920}.

\bibitem[{\citenamefont{Ruffert et~al.}(1996)\citenamefont{Ruffert, Janka, and
  Schaefer}}]{Ruffert:1995fs}
\bibinfo{author}{\bibfnamefont{M.~H.} \bibnamefont{Ruffert}},
  \bibinfo{author}{\bibfnamefont{H.~T.} \bibnamefont{Janka}}, \bibnamefont{and}
  \bibinfo{author}{\bibfnamefont{G.}~\bibnamefont{Schaefer}},
  \bibinfo{journal}{Astron. Astrophys.} \textbf{\bibinfo{volume}{311}},
  \bibinfo{pages}{532} (\bibinfo{year}{1996}), \eprint{astro-ph/9509006}.

\bibitem[{\citenamefont{Rosswog and Liebendoerfer}(2003)}]{Rosswog:2003rv}
\bibinfo{author}{\bibfnamefont{S.}~\bibnamefont{Rosswog}} \bibnamefont{and}
  \bibinfo{author}{\bibfnamefont{M.}~\bibnamefont{Liebendoerfer}},
  \bibinfo{journal}{Mon. Not. Roy. Astron. Soc.}
  \textbf{\bibinfo{volume}{342}}, \bibinfo{pages}{673} (\bibinfo{year}{2003}),
  \eprint{astro-ph/0302301}.

\bibitem[{\citenamefont{Loffler et~al.}(2012)}]{Loffler:2011ay}
\bibinfo{author}{\bibfnamefont{F.}~\bibnamefont{Loffler}} \bibnamefont{et~al.},
  \bibinfo{journal}{Class. Quant. Grav.} \textbf{\bibinfo{volume}{29}},
  \bibinfo{pages}{115001} (\bibinfo{year}{2012}), \eprint{1111.3344}.

\bibitem[{\citenamefont{Schnetter et~al.}(2004)\citenamefont{Schnetter, Hawley,
  and Hawke}}]{Schnetter:2003rb}
\bibinfo{author}{\bibfnamefont{E.}~\bibnamefont{Schnetter}},
  \bibinfo{author}{\bibfnamefont{S.~H.} \bibnamefont{Hawley}},
  \bibnamefont{and} \bibinfo{author}{\bibfnamefont{I.}~\bibnamefont{Hawke}},
  \bibinfo{journal}{Class. Quant. Grav.} \textbf{\bibinfo{volume}{21}},
  \bibinfo{pages}{1465} (\bibinfo{year}{2004}), \eprint{gr-qc/0310042}.

\bibitem[{\citenamefont{Kiuchi et~al.}(2017)\citenamefont{Kiuchi, Kawaguchi,
  Kyutoku, Sekiguchi, Shibata, and Taniguchi}}]{Kiuchi:2017pte}
\bibinfo{author}{\bibfnamefont{K.}~\bibnamefont{Kiuchi}},
  \bibinfo{author}{\bibfnamefont{K.}~\bibnamefont{Kawaguchi}},
  \bibinfo{author}{\bibfnamefont{K.}~\bibnamefont{Kyutoku}},
  \bibinfo{author}{\bibfnamefont{Y.}~\bibnamefont{Sekiguchi}},
  \bibinfo{author}{\bibfnamefont{M.}~\bibnamefont{Shibata}}, \bibnamefont{and}
  \bibinfo{author}{\bibfnamefont{K.}~\bibnamefont{Taniguchi}},
  \bibinfo{journal}{Phys. Rev. D} \textbf{\bibinfo{volume}{96}},
  \bibinfo{pages}{084060} (\bibinfo{year}{2017}), \eprint{1708.08926}.

\bibitem[{\citenamefont{Foucart et~al.}(2019)}]{Foucart:2018lhe}
\bibinfo{author}{\bibfnamefont{F.}~\bibnamefont{Foucart}} \bibnamefont{et~al.},
  \bibinfo{journal}{Phys. Rev. D} \textbf{\bibinfo{volume}{99}},
  \bibinfo{pages}{044008} (\bibinfo{year}{2019}), \eprint{1812.06988}.

\bibitem[{\citenamefont{Kiuchi et~al.}(2020)\citenamefont{Kiuchi, Kawaguchi,
  Kyutoku, Sekiguchi, and Shibata}}]{Kiuchi:2019kzt}
\bibinfo{author}{\bibfnamefont{K.}~\bibnamefont{Kiuchi}},
  \bibinfo{author}{\bibfnamefont{K.}~\bibnamefont{Kawaguchi}},
  \bibinfo{author}{\bibfnamefont{K.}~\bibnamefont{Kyutoku}},
  \bibinfo{author}{\bibfnamefont{Y.}~\bibnamefont{Sekiguchi}},
  \bibnamefont{and} \bibinfo{author}{\bibfnamefont{M.}~\bibnamefont{Shibata}},
  \bibinfo{journal}{Phys. Rev. D} \textbf{\bibinfo{volume}{101}},
  \bibinfo{pages}{084006} (\bibinfo{year}{2020}), \eprint{1907.03790}.

\end{thebibliography}

\pagebreak
\widetext
\begin{center}
\textbf{\large Supplemental material for:\\ Degeneracy in the inference of phase transitions in the neutron star equation of state from gravitational wave data}
\end{center}
\FloatBarrier
\setcounter{equation}{0}
\setcounter{figure}{0}
\setcounter{table}{0}
\setcounter{page}{1}
\makeatletter
\renewcommand{\theequation}{S\arabic{equation}}
\renewcommand{\thefigure}{S\arabic{figure}}

\section{Bayesian inference with additional priors}

\begin{figure*} \centering
  \includegraphics[width=\textwidth]{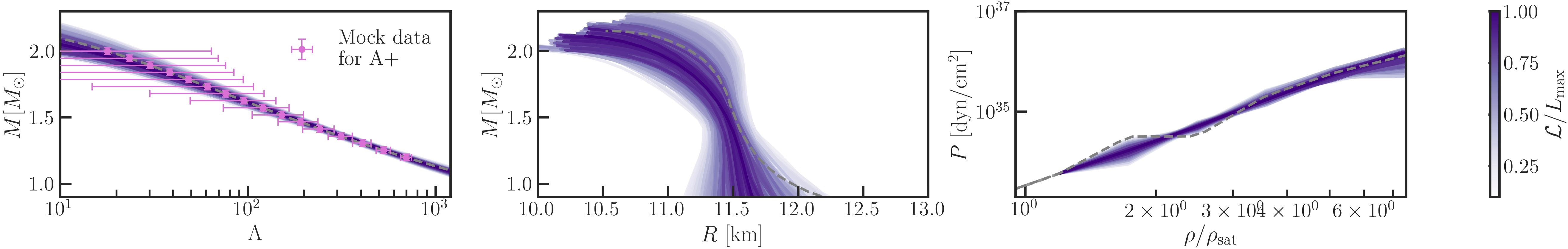}\\
    \includegraphics[width=\textwidth]{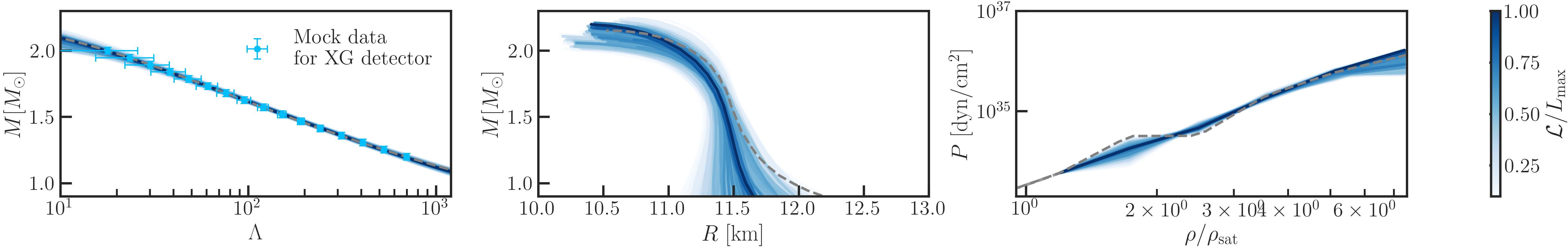}
  \caption{\label{fig:inference_reg} Bayesian inference of the equation of state from mock data that was generated assuming Gaussian errors in tidal deformability, $\Lambda$, for a series of GW170817-like events observed with the sensitivity of LIGO in the A+ configuration ($\sigma_{\Lambda}$=46; top panel); and for the proposed XG detector Cosmic Explorer ($\sigma_{\Lambda}$=8; bottom panel). In both cases, the mock data are generated from the same EoS (gray dashed lines). From left to right, we show: the most likely tidal deformability curves, mass-radius curves, and EoSs inferred in our Bayesian inference. The curves shown are randomly sampled from the 68\% confidence interval and colored according to their posteriors, relative to the most likely solution.} 
\end{figure*}

We perform an additional set of Bayesian inferences for the mock data used in Fig. 1 of the main Letter,
with a regularizer applied to weakly penalize variations in the sound speed. The regularizer
is given by
\begin{equation}
\label{eq:reg}
\xi = \exp{\frac{\left(\partial^2 \ln P/\partial \ln \rho^2\right)^2}{2\lambda^2}}
\end{equation}
where $P$ is the pressure, $\rho$ is the mass density, and $\lambda$ is a characteristic scale. We conservatively set $\lambda=8$, following Ref.~\cite{Raithel:2017ity}, who showed that the second logarithmic derivatives for a wide range of tabulated EoSs, $(\partial^2 \ln P/\partial \ln \rho^2)$, are typically $\lesssim2$. The choice of $\lambda=8$ thus applies a weak prior to penalize extreme density-variations in the EoS. The likelihood for each EoS drawn during the Markov Chain Monte Carlo fitting procedure is multiplied by this regularizer. The details of the inference are otherwise identical to what was performed in the main Letter and follow the methodology of \cite{Raithel:2017ity}. We show the results of these inferences in Fig.~\ref{fig:inference_reg}.

As expected, this regularizer has significantly restricted 
the 68\% uncertainty bands for the inferred EoSs, compared to the 
assumption of purely uniform priors in pressure (as in Fig.~1 of the
main Letter). For example, $R_{1.4}$ is now constrained to within 420~m (390~m)
at 68\% confidence for the A+ (CE) mock data, while the 68\% error band 
in the pressure at 1.7$\rns$ now spans a factor of 3.6$\times$ (3.2$\times$).

However, Fig.~\ref{fig:inference_reg} also demonstrates that -- \textit{even with the sensitivity of the 16 mock Cosmic Explorer measurements} -- the choice of this weak prior acts to select the incorrect EoS, as can be seen in the mismatch between the most-likely inferred EoS (in dark blue) and the EoS that was used to generate the mock data (in gray). In other words, there is a continuous range of EoS parameter space that can fit the mock data comparably well, even in the limit of high-quality data observed across a wide range of masses. This is a direct consequence of the \dop degeneracy. As a result, the choice of priors in such inferences will be very important for some regions of the EoS parameter space. 
\FloatBarrier

\begin{figure*} \centering
  \includegraphics[width=\textwidth]{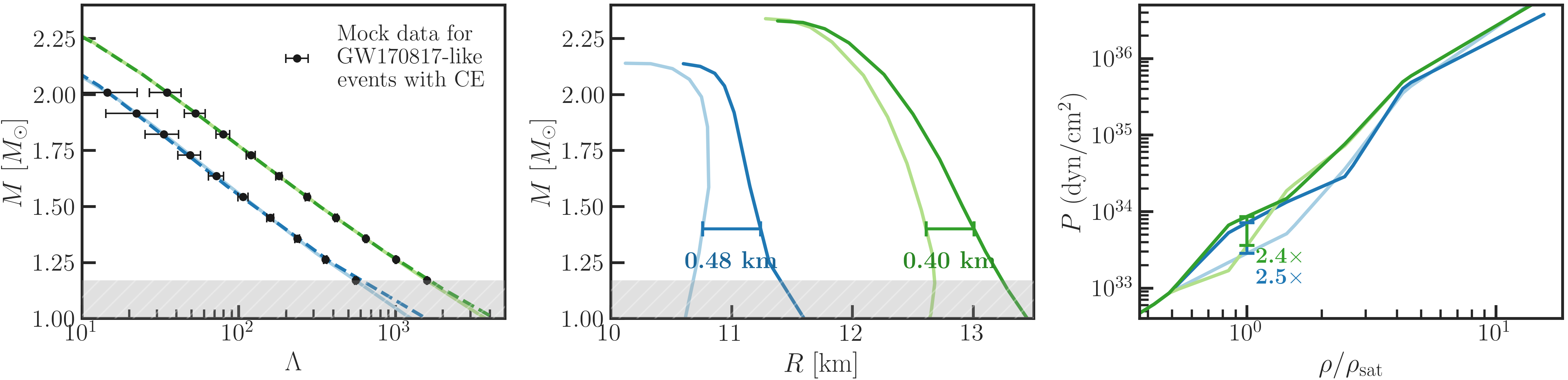}
  \caption{\label{fig:MR} Two examples pairs of \dop EoSs, colored in blue
  and green. The crust EoS is taken to be the zero-temperature, $\beta$-equilibrium slice of SFHo \cite{Steiner:2012rk}, and is assumed to 0.5$\rns$. 
  From left to right, we show: the neutron star tidal deformability $\Lambda$ as a function
  of mass $M$, the mass-radius relation, and the pressure-density function.
  The left panel also includes a mock sample of tidal deformability measurements
  for a series of GW170817-strength events observed with Cosmic Explorer
  (implying a 68\% measurement uncertainty of $\sigma_{\Lambda}=8$ \cite{Carson:2019rjx}).
    The gray bands indicate
  the lowest observed neutron star mass of 1.17~$\Ms$.
  The \dops differ by $\lesssim30$ in their tidal deformabilities across
  the range of observed neutron star masses, but
  differ by 0.4-0.5~km in the radius of a 1.4~$\Ms$ neutron star and
   by $\sim$2.5 times in the pressure at the nuclear saturation density.}
\end{figure*}

\section{Equations of state for the simulated pairs of \dops}
\label{app:eos}

We construct the zero-temperature, $\beta$-equilibrium equation of state
(EoS) using the piecewise polytropic (PWP) framework of
\cite{Read:2008iy,Ozel:2009da,Raithel:2016bux}. In particular, we use five
piecewise polytropic segments which are spaced uniformly in the logarithm
of the density between $\rho_0$ and $7.4\rns$, and use a tabulated crust
EoS at densities below $\rho_0$. In the PWP formulation, the pressure $P$
between two fiducial densities $\rho_{i-1}$ and $\rho_i$ is given by
\begin{equation}
P(\rho) = K_i \rho^{\Gamma_i}, \quad \rho_{i-1} < \rho < \rho_i
\end{equation}
where $\rho$ is the density, the polytropic constant, $K_i$, is determined
by requiring continuity between adjacent polytropic segments, according to
\begin{equation}
K_i = \frac{P_{i-1}}{\rho_{i-1}^{\Gamma_i}} = \frac{P_i}{\rho_i^{\Gamma_i}},
\end{equation}
and the polytropic index, $\Gamma_i$, is given by
\begin{equation}
    \Gamma_i \equiv \frac{\partial \ln P}{\partial \ln \rho} = \frac{\ln\left( P_i / P_{i-1} \right)}{\ln \left( \rho_i/\rho_{i-1} \right)}.
\end{equation}

For the two pairs of \dop EoSs shown in Fig.~S2, we start our parametrization at $\rho_0=0.5\rns$. For the crust EoS
at $\rho<\rho_0$, we use the SFHo EoS \cite{Steiner:2012rk}. In order to
ensure continuity in the EoS, we fix the pressure at $\rho_0$ to that of
SFHo. We then vary the pressures at higher densities to construct examples
of tidal deformability degeneracy, while enforcing a set of minimal
physical constraints, namely that the sound speed remain sub-luminal, that
the star remain hydrostatically stable, and that the EoS be able to support
massive ($2~\Ms$) neutron stars.

We report the pressures that uniquely characterize the PWP parametrization
for the four \dops EoSs from Fig.~S2 in Table~\ref{table:PWP}.

\begin{table*}
\centering
\begin{tabularx}{0.9\textwidth}{@{\extracolsep{\fill}}llllllll}
\hline \hline
 $R_{1.4}$ [km]   & $\Lambda_{1.4}$ & $P(0.86\rns)$ & $P(1.47\rns)$ & $P(2.52\rns)$ & $P(4.32\rns)$ & $P(7.40\rns)$ &  \\
\hline   
10.77  &  193  &  2.22e+33 & 5.46e+33 & 3.90e+34 & 3.76e+35 & 1.30e+36 \\
11.23  &  193  & 5.44e+33 & 1.36e+34 & 3.12e+34 & 4.25e+35 & 1.09e+36 \\
\hline
12.64  &  522  &  1.82e+33 & 1.97e+34 & 7.75e+34 & 5.21e+35 & 1.50e+36 \\
13.01  &  522  &  6.81e+33 & 1.56e+34 & 8.15e+34 & 5.22e+35 & 1.50e+36 \\
\hline 
\end{tabularx}
\caption{Model parameters for four example \dop EoSs. The first column
reports the radius of a 1.4~$\Ms$ neutron star; the second column reports
the tidal deformability of a 1.4~$\Ms$ star; and the remaining columns
report the pressures at each fiducial density in our parametrization.  All
pressures are in units of dyn/cm$^2$. }
  \label{table:PWP}
\end{table*}

For use in our numerical merger simulations, we extend these
zero-temperature, $\beta$-equilibrium EoSs to finite-temperatures and
arbitrary compositions using the framework of \cite{Raithel:2019gws}, which was recently
validated in the context of merger simulations in \cite{Raithel:2022nab}. The
finite-temperature prescription is based on a two-parameter model of the
particle effective mass, for which we adopt the parameters
$n_0=0.12$~fm$^{-3}$ and $\alpha=0.8$, which correspond to an intermediate
set of parameters compared to values that are fit to a sample of nine
existing finite-temperature EoS tables (as reported in
\cite{Raithel:2019gws}). The
extension to arbitrary proton fraction is based on a parametrization of the
nuclear symmetry energy. We use symmetry energy parameters $S_0=32$~MeV for
all four EoSs while the leading-order slope parameter $L$ is calculated in
terms of the pressure at $\rns$, according to eq.~(6) of \cite{Most:2021ktk}.
This
corresponds to $L=40$~MeV and 100~MeV for the $R_{1.4}=10.8$~km and 11.2~km
models, respectively, and $L=50$~MeV and 120~MeV for the $R_{1.4}=12.6$~km
and 13~km models. At densities below $\rns$, we smoothly connect to the
full SFHo table to describe the low-density EoS, using the free-energy
matching scheme of \cite{Schneider:2017tfi}. The implementation details for constructing the
full, finite-temperature versions of these EoS tables are otherwise
identical to the procedure described in \cite{Most:2021ktk}. 

\section{Binary tidal deformability}
\label{sec:binary}
\begin{figure}[!ht]
\centering
\includegraphics[width=0.9\textwidth]{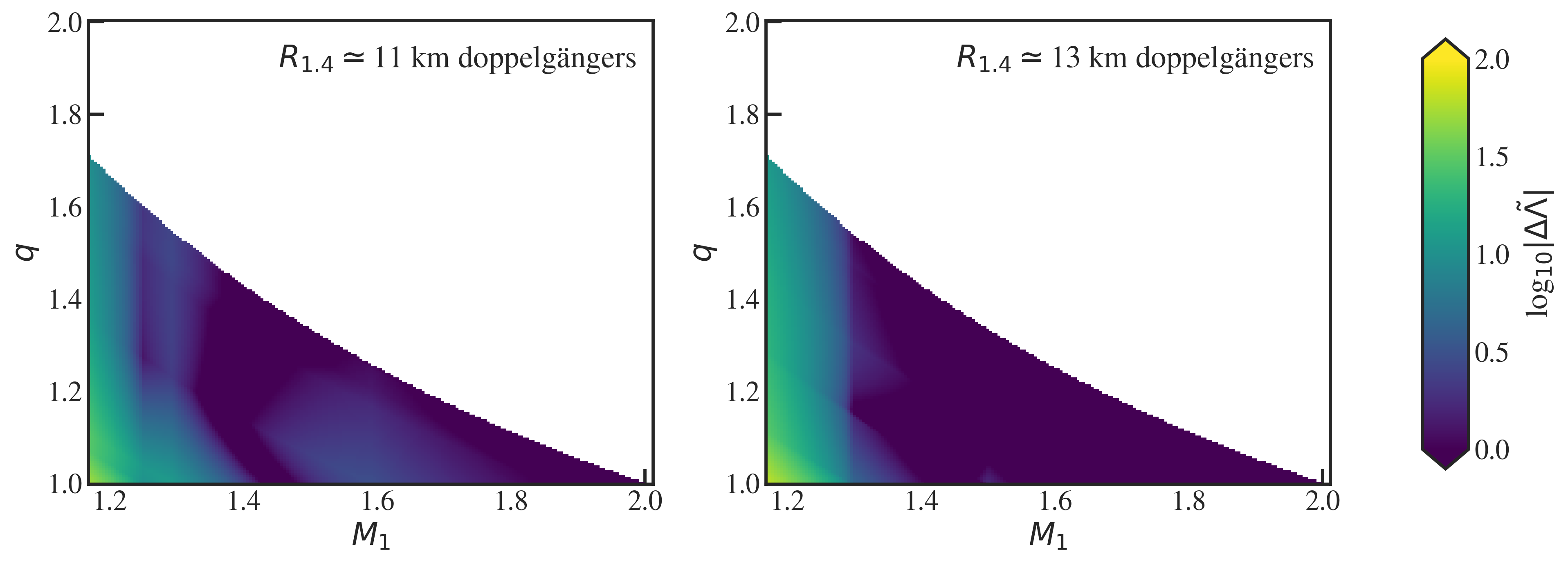}
\caption{\label{fig:dLambda} Contours showing the difference in binary tidal deformability,
$|\Delta\widetilde{\Lambda}|$, as a function of the primary mass, $m_1$, and the mass ratio, $q$,
for the softer pairs of \dop EoSs ($R_{1.4}\simeq11$~km) from Fig.~S2 in the left panel,
and the stiffer pair ($R_{1.4}\simeq13$~km) in the right panel.}
\end{figure}

The inspiral gravitational waves are most sensitive not to the tidal deformabilities of the individual neutron stars ($\Lambda_1$ and $\Lambda_2$), but rather to the binary tidal deformability, $\tilde{\Lambda}$, which is a mass-weighted average of the two, defined according to
\begin{equation}
\label{eq:leff}
    \widetilde{\Lambda} = \frac{16}{13} \frac{(m_1 + 12 m_1)m_1^4\Lambda_1 + (m_2+12 m_1)m_2^4 \Lambda_2}{ (m_1 + m_2)^5}
\end{equation}
where $m_{1,2}$ are the component masses.
We calculate the binary tidal deformability for each of the simulated pairs of \dops from  Fig.~S2, and we show the differences between each pair of \dops for a range of masses and mass ratios in Fig.~\ref{fig:dLambda}.
 As expected given the similarity between the tidal deformabilities of these models, the differences in $\tilde{\Lambda}$ are also very small, and are $\lesssim10$ for most binary masses.

\section{Numerical relativity simulations of \dops}
\label{sec:sims}
For our numerical relativity simulation of the \dop models, we use the same
setup as in \cite{Most:2021ktk}. In fact the simulation results for the
$R_{1.4}\simeq11\,\rm km$ models  have previously been presented there.
In this section, we summarize the main features of the simulations.
We solve the coupled Einstein-hydrodynamics system using the Frankfurt-/IllinoisGRMHD code (\texttt{FIL})
code \cite{Most:2019kfe,Most:2018eaw,Etienne:2015cea}. This solves the
general-relativistic (magneto-)hydrodynamics system
\cite{Duez:2005sf,Shibata:2005gp} together with the Z4c formulation of
Einsteins equations \cite{Hilditch:2012fp,Bernuzzi:2009ex}. In addition, we
include a neutrino leakage scheme \cite{Ruffert:1995fs,Rosswog:2003rv}, and
evolve the system for vanishing magnetic fields. The computational
infrastracture is provided by the \texttt{Einstein Toolkit}
\cite{Loffler:2011ay} infrastructure.
We impose reflection symmetry across the orbital plane, and adopt a
finest-level
resolution of $262\,\rm m$, with 8 levels of fixed mesh refinement
\cite{Schnetter:2003rb}. The simulations are performed for about
$10-15\,\rm ms$ after the merger, sufficient to extract the $f_2$ peak
frequency of the post-merger spectrum.

\begin{figure}[!ht]
\centering
\includegraphics[width=0.9\textwidth]{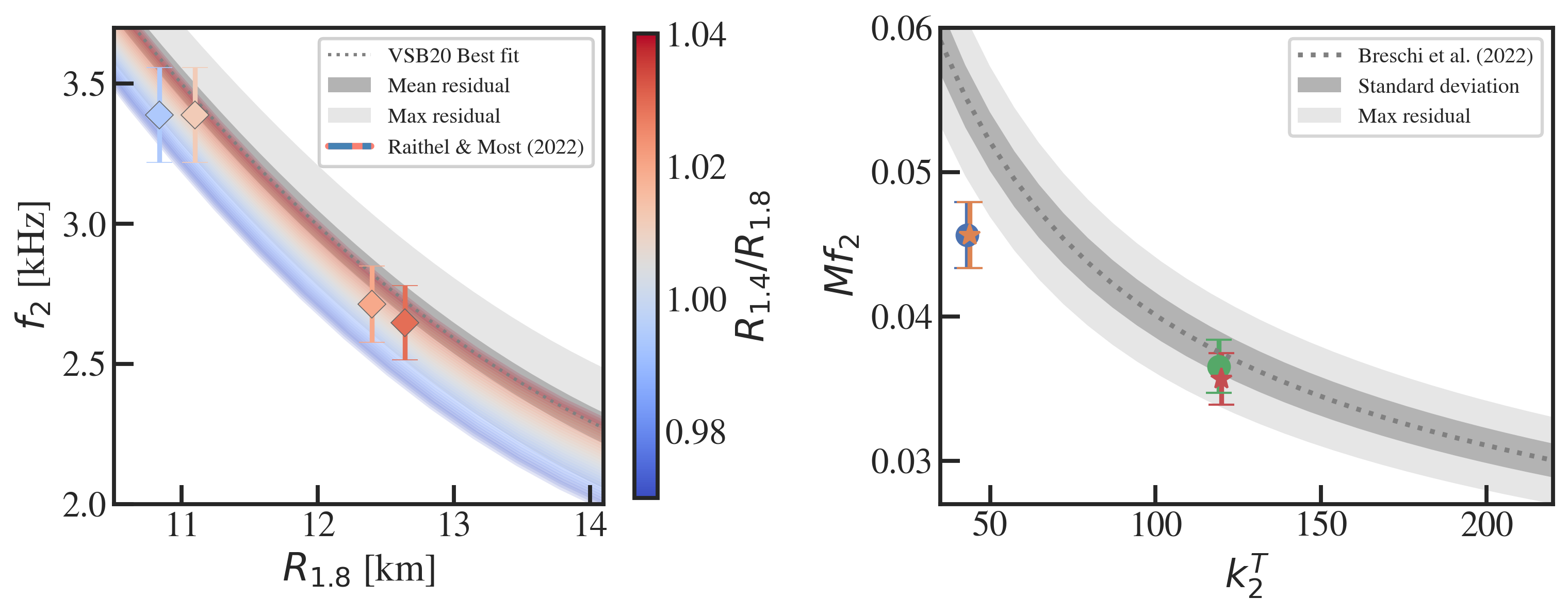}
\caption{\label{fig:f2} Quasi-universal relations between the peak frequency of the post-merger  
gravitational wave signal, $f_2$, and properties of the neutron star EoS. Left:
correlation between $f_2$ and the radius of a 1.8$\Ms$ neutron star from \cite{Vretinaris:2019spn} in gray,
with the colored lines showing the two-parameter correlation with $R_{1.8}$ and the slope of the mass-radius relation
\cite{Raithel:2022orm}. Right: correlation between $f_2$ and the binary tidal parameter, $k_2^T$,
from \cite{Breschi:2022xnc}. The markers correspond to the four merger simulations performed
for the two example pairs of \dop EoSs shown in Fig.~\ref{fig:MR}. 
We have assumed a $10\%$ error on $f_2$ consistent with \cite{Breschi:2019srl}.} 
\end{figure}

We summarize the peak frequencies of the
post-merger GW emission for each of the four simulations in
Fig.~\ref{fig:f2}, where we compare these results against two existing
quasi-universal relations \cite{Vretinaris:2019spn, Breschi:2022xnc},  as well as one with a proposed correction that depends
on the mass-radius slope, appropriate for the EoS models considered here
\cite{Raithel:2022orm}. We conservatively estimate the numerical uncertainty of $f_2$
to be at the 10\% level \cite{Breschi:2019srl}. Starting with the correlation between $f_2$ and the radius
  for a $1.8 M_\odot$ star, $R_{1.8}$, we find that (to within current
  numerical and systematic uncertainties) the \dop models considered here
  do not violate the existing relations of
  \cite{Vretinaris:2019spn,Raithel:2022orm} shown in the left panel of Fig. \ref{fig:f2}.
  Concerning the tidal coupling constant $\kappa_2^{T}$ (as defined in eq.~6 of \cite{Breschi:2022xnc}),
the relation with $f_2$ is also approximately consistent with
the existing relations proposed by \cite{Breschi:2022xnc}, to within the
numerical errors of the calculations. We note that the small disagreement between
the computed frequencies at small $\kappa_2^T$ and the universal relations of
\cite{Breschi:2022xnc} is likely a result of numerical uncertainty and limited
calibration in this regime in terms of EoS coverage
\cite{Kiuchi:2017pte,Foucart:2018lhe,Kiuchi:2019kzt,Breschi:2019srl,Breschi:2022xnc},
and may also be affected by uncertainties in
finite-temperature
\cite{Bauswein:2010dn,Figura:2020fkj,Raithel:2021hye} and neutrino physics
\cite{Alford:2017rxf,Most:2021zvc,Radice:2021jtw,Most:2022yhe}.

\end{document}